\documentclass{article}%
\usepackage{placeins}
\usepackage{tikz}
\usepackage{MathShorthand}
\usepackage{amsmath}
\usepackage{amsfonts}
\usepackage{amssymb}
\usepackage{graphicx}
\usepackage{caption}
\usepackage{xcolor}
\usepackage{subcaption}
\usepackage{fullpage}
\usepackage{amsthm}
\usepackage{wasysym}
\usepackage{epstopdf}
\usepackage{comment}%
\providecommand{\U}[1]{\protect\rule{.1in}{.1in}}
\usetikzlibrary{calc}

\def\XXint#1#2#3{{\setbox0=\hbox{$#1{#2#3}{\int}$}
\vcenter{\hbox{$#2#3$}}\kern-.5\wd0}}

\begin{document}

\title{Multi-vortex crystal lattices in Bose-Einstein Condensates with a rotating trap}
\author{Shuangquan Xie, Panayotis G. Kevrekidis and Theodore Kolokolnikov}
\maketitle

\begin{abstract}
We consider vortex dynamics in the context of Bose-Einstein Condensates (BEC)
with a rotating trap, with or without anisotropy. Starting with the
Gross-Pitaevskii (GP) partial differential equation (PDE), we derive a novel
reduced system of ordinary differential equations (ODEs) that describes stable
configurations of multiple co-rotating vortices (vortex crystals). This 
description is found to be quite accurate \textit{quantitatively} especially
in the case of multiple vortices. In the limit of many vortices, BECs are
known to form vortex crystal structures, whereby vortices tend to arrange
themselves in a hexagonal-like spatial configuration. Using our asymptotic
reduction, we derive the effective vortex crystal density and its radius. We
also obtain an asymptotic 
estimate for the maximum number of vortices as a function of
rotation rate. We extend considerations to the anisotropic trap case,
confirming that a pair of vortices lying on the long (short) axis is linearly
stable (unstable), corroborating the ODE reduction results with full PDE
simulations. We then further investigate the many-vortex limit in the case
of strong anisotropic potential. In this limit, the vortices tend to align
themselves along the long axis, and we compute the effective one-dimensional
vortex density, as well as the maximum admissible number of vortices. Detailed
numerical simulations of the GP\ equation are used to confirm our analytical predictions.

\end{abstract}

\section{Introduction}

Theoretical and experimental studies on vortices in rotating Bose-Einstein
Condensates (BEC) have attracted great interest in the past 20 years, see,
e.g. \cite{castin1999bose}, the review \cite{fetter2009rotating} and the
monographs \cite{aftalion2007vortices,siambook} where extensive lists of
references can be found. In most of the theoretical research, the
Gross-Pitayevskii equation (GPE) model has served to study the emergence and
dynamics of vortices. As an approximation of the quantum mechanical many-body
problem at zero temperature, Gross-Pitaevskii theory was rigorously
established in \cite{lieb2001rigorous} for the non-rotating case and in
\cite{lieb2006derivation} for rotating systems.

One of the most interesting features observed experimentally is that when the
angular speed gets larger, vortices are spontaneously
nucleating~\cite{fetter2009rotating}, since their presence minimizes the
system's free energy. As the frequency of rotation is increased, the number of
vortices increases and they eventually arrange themselves in a hexagonal
lattice-like pattern around the center of the condensate
\cite{abo2001observation,madison2000vortex}. It is natural to explore the
mechanism of this behavior mathematically. Under the framework of GP theory,
the critical angular velocity was rigorously computed in
\cite{serfaty2001model,ignat2006critical} and the distribution of the first
few vortices to appear in the condensate was studied in
\cite{ignat2006energy}. Another striking observation in experiments is that
the vortex lattice seems to be nearly homogeneous even when the matter density
profile of the condensate imposed by the trap is not homogeneous
\cite{bretin2004fast, schweikhard2004rapidly}. The relation between the matter
density and the vortex density has been formulated in
\cite{sheehy2004vortices, sheehy2004vortex}. However,
Ref.~\cite{correggi2013inhomogeneous} argues that the vortex distribution is
strongly inhomogeneous close to the critical speed for vortex nucleation
and gradually homogenizes
when the rotation speed is increased. The study of both such vortex lattices
and also of small scale vortex clusters~\cite{navarro2013dynamics,zampetaki},
thus, remains an active topic of both theoretical and experimental investigation.

\begin{figure}[tb]
{\centering
\includegraphics[width=0.49\textwidth]{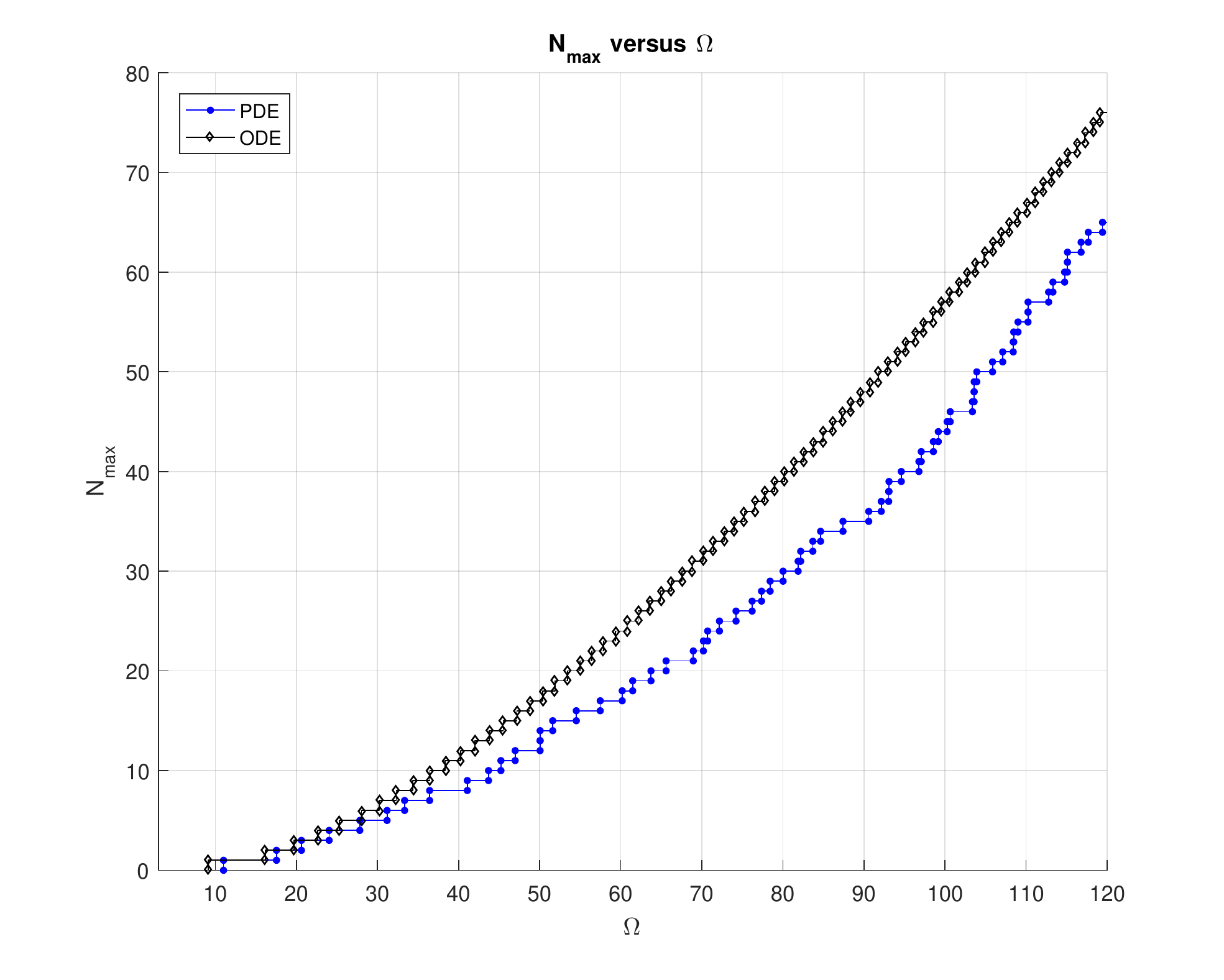}
\includegraphics[width=0.49\textwidth]{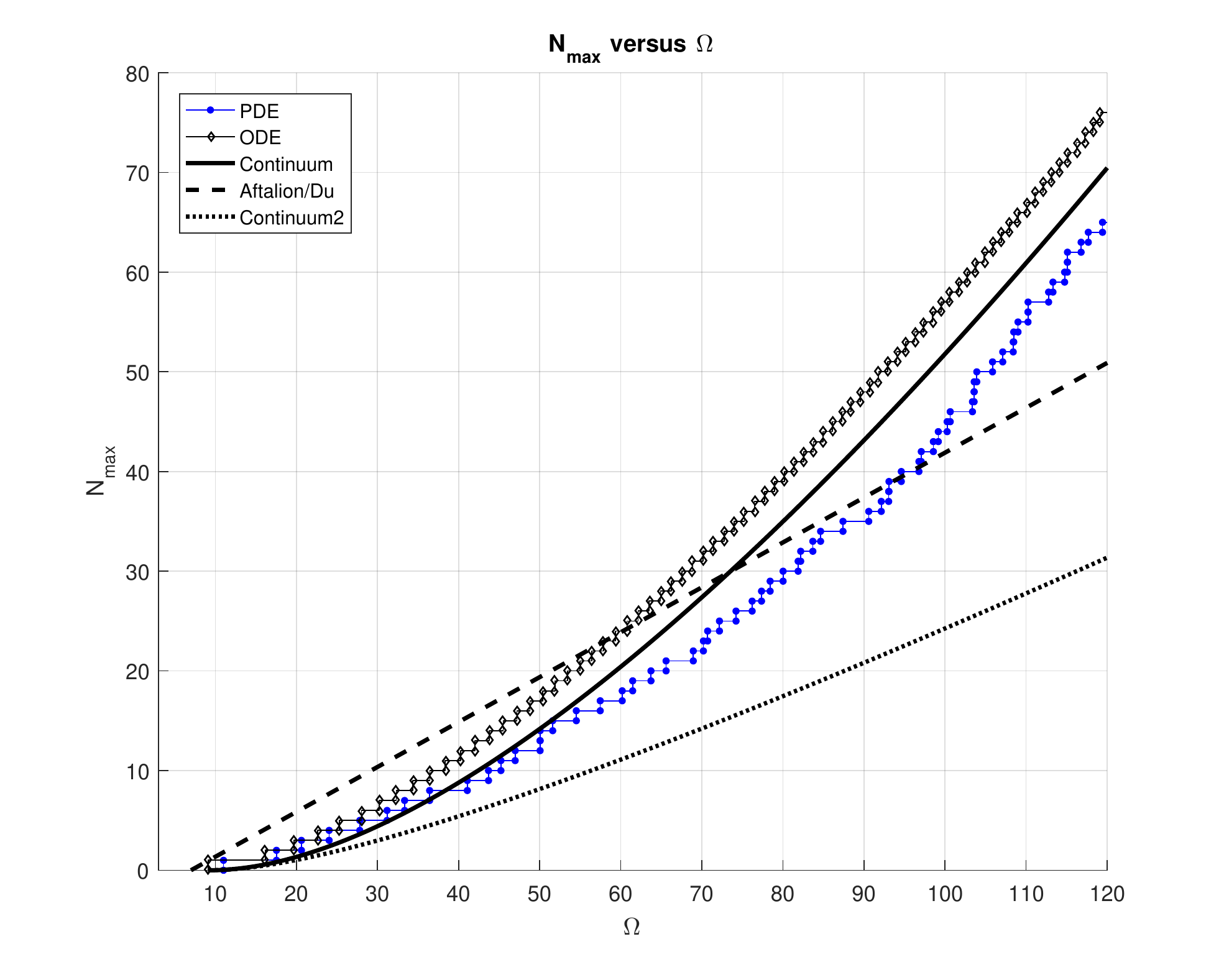} }
\par
\setlength{\unitlength}{0.05\textwidth} \tikz[overlay,remember picture] {
\put(1.3,2){\includegraphics[width=0.1\textwidth]{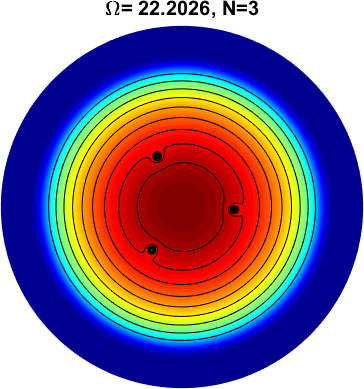}}
\put(3.0,3.3){\includegraphics[width=0.1\textwidth]{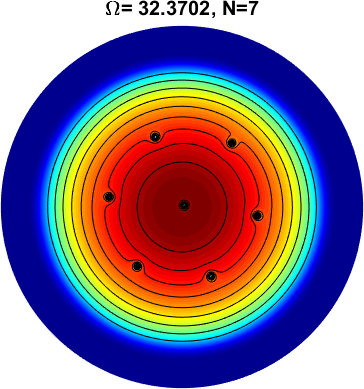}}
\put(5.8,1.5){\includegraphics[width=0.1\textwidth]{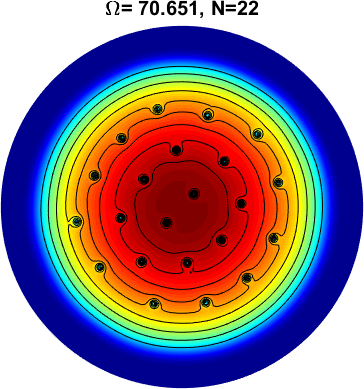}}
\put(7.5,3.0){\includegraphics[width=0.1\textwidth]{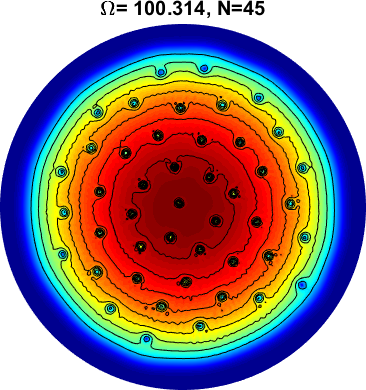}}
\put(6.0,6.0){\includegraphics[width=0.1\textwidth]{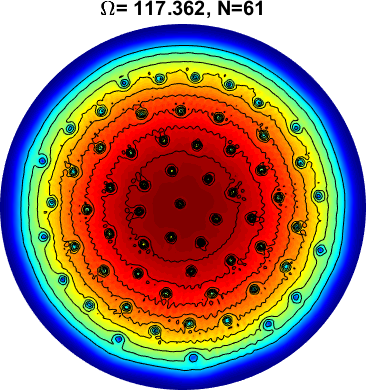}}
\put(5,0){ $\mbox{(a)}$}
\node[text width=0cm] at (4,0.2) {(a)};
\node[text width=0cm] at (12,0.2) {(b)};
\draw [-latex, thick, red] (7.5,0.6) -- (1.3, 0.6);
\draw [-latex] (2.03, 1.65) -- (2.05,1.38);
\draw [-latex] (3.05, 2.7) -- (2.70,1.58);
\draw [-latex]  (4.88,2.45) -- (4.68,2.65);
\draw [-latex] (6.45,3.85) -- (6.25,4.05);
\draw [-latex] (6.3,5.80) -- (7.05,5.15);
} \caption{ (a) Maximum number of vortices as a function of $\Omega$.
\textquotedblleft PDE\textquotedblright\ denotes the full PDE simulation of
(\ref{gp}) in the overdamped regime ($\kappa=0,\gamma=1$). We start with
$\Omega=125$ and an initial configuration of 80 vortices. Then $\Omega$ is
decreased very slowly in time according to the formula $\Omega=125-10^{-4}t$
(indicated by a red arrow). Other parameters are $b=1,\varepsilon=0.01$. We
count the number of vortices at each value of $\Omega$, and this is what is
plotted. \textquotedblleft ODE\textquotedblright\ denotes the simulation of
the reduced ODE system (\ref{reduced}), with the same parameters as the
PDE. See
remarks following Eq. (\ref{Constrain_iso}) for further details of PDE/ODE
simulations. Snapshots show steady states of the PDE for several values of
$\Omega$. (b) Comparison to previous results. \textquotedblleft
PDE\textquotedblright\ and \textquotedblleft ODE\textquotedblright\ are the
same as in (a). \textquotedblleft Continuum\textquotedblright\ refers to
Eq.~(\ref{Nmax}). \textquotedblleft Aftalion/Du\textquotedblright\ is the
Eq.~(\ref{NmaxAD}) originally derived in \cite{aftalion2001vortices}. Finally,
\textquotedblleft Continuum2\textquotedblright\ represents Eq.~(\ref{Nmax2})
first derived in \cite{kolokolnikov2014tale}.}%
\label{fig:Nmax}%
\end{figure}
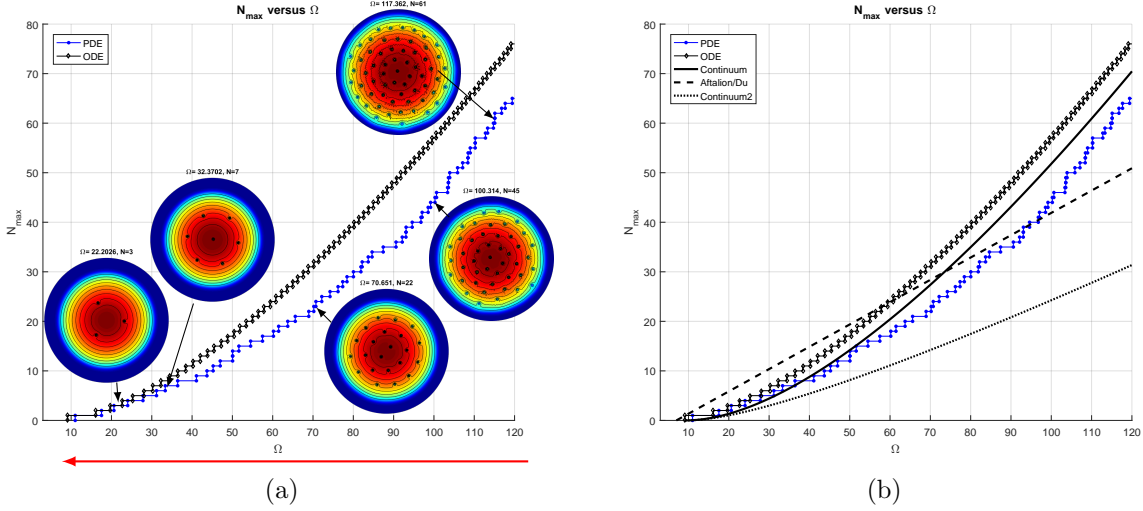

In this paper, we use asymptotic techniques following
\cite{weinan1994dynamics} to derive a novel set of equations which describe
the distribution of vortex lattices in rotating BEC~\footnote{Admittedly,
there are numerous other techniques that enable the derivation of such vortex
equations, including the use of conservation laws~\cite{jerrard}, as well as
of variational principles~\cite{variational}. Here, we focus on the asymptotic
techniques of~\cite{weinan1994dynamics}.}. The equations we derive are valid
for both the isotropic and the anisotropic case. We then use the new equations
to study the following important limits:

\begin{itemize}
\item \textbf{Many-vortex limit, isotropic trap: }This is the limit where
vortex crystals are observed. By taking a continuum limit of the effective
equations of motion, we consider the equilibrium of the effective density of
the vortex crystals, as well as the size of the lattice. In addition, this
computation yields an asymptotic estimate for
the \emph{maximum} number of vortices that can form stable
lattice configurations, as a function of rotation speed. This is illustrated
in Figure \ref{fig:Nmax}.

\item \textbf{High anisotropy many-vortex regime:} When the anisotropy is
sufficiently high, the vortices tend to align along the longer axis of the
trap; see Figure \ref{fig:anisotropy}. This constitutes the energetically
favorable configuration. In this limit, we compute the one-dimensional density
of the resulting vortex configuration by using techniques involving the
Chebyshev polynomials. As in the isotropic case, this leads to an expression
relating the maximum number of vortices in a stable configuration and other
problem parameters such as the anisotropy and the rotation rate.
\end{itemize}

We validate our results by a direct comparison of the reduced particle ODEs
with the full numerical solution. The PDE system is simulated using the
finite-element package FlexPDE6~\cite{flex}. FlexPDE6 uses adaptive mesh in
space, and adaptive time stepping. This is particularly useful for computing
vortex solutions which are localized in space. In our computations
we used up
to 40000 nodes with global error tolerances up to $10^{-4}$. To validate the
numerics we verified that doubling mesh size and error tolerances did not
affect the overall results.

Our starting point is the Gross-Pitaevskii (GP) equation with an inhomogeneous
rotating trap in two dimensions given by\bes\label{gp}%
\begin{equation}
(\gamma-\kappa i)w_{t}=\Delta w+\frac{1}{\varepsilon^{2}}\left(
V(x)-|w|^{2}\right)  w+i\Omega\left(  x_{2}w_{x_{1}}-x_{1}w_{x_{2}}\right)  .
\label{GBP}%
\end{equation}
The parameter $\varepsilon$ is assumed to be small, which corresponds to the
large chemical potential (also known as semiclassical~\cite{siambook}) limit.
$\Omega$ is the rotation rate, and $V(x)$ is the trap potential. We consider
the general anisotropic parabolic potential \footnote{{Note that it is easy to
extend to the result to the more general case: $V(x)=1-b_{1}^{2}x_{1}%
^{2}-b_{2}^{2} x_{2}^{2}$. In fact, we could just rescale $\hat{t}=\frac{t}{b_1^2},~\hat{x}_{1}=b_{1} x_{1},~\hat{x}_{2}=b_{1} x_{2}$ and define
$b=\frac{b_{2}}{b_{1}},~\hat{\varepsilon}=b_{1} \varepsilon,~\hat{\Omega
}=\frac{\Omega}{b_{1}^{2}} $ so that the PDE (\ref{gp}) becomes:
\[
(\gamma-\kappa i)w_{\hat{t}}=\Delta w+\frac{1}{\hat{\varepsilon}^{2}}\left(
1-\hat{x}_{1}^{2}-b^{2}\hat{x}_{2}^{2}-|w|^{2}\right)  w+i\hat{\Omega} \left(
\hat{x}_{2} w_{\hat{x}_{1}}-\hat{x}_{1}w_{\hat{x}_{2}}\right) .
\]
}} $V(x),$%
\begin{equation}
V(x)=1-x_{1}^{2}-b^{2}x_{2}^{2} \label{elltrap}%
\end{equation}
\ees The parameter $b$ represents the strength of the anisotropy, with
the isotropic trap limit corresponding to $b=1.$ Here, we use the notation $V=1-
\tilde{V}$ where $\tilde{V}$ represents the customary confining parabolic
trap. Finally, the ratio $\gamma/\kappa$ represents the finite temperature
effects; see the relevant discussion
in~\cite{tsubota2002vortex, penckwitt2002nucleation}. For the purposes of
numerical simulations, we mostly work in the the overdamped regime
$\gamma/\kappa\rightarrow\infty$, sometimes referred to as imaginary time
integration \cite{penckwitt2002nucleation, feder1999vortex}. While the
equilibrium vortex lattice state is independent of $\gamma$, numerical
simulations are easier to perform in the overdamped regime. I.e., Our aim from
the point of view of numerical computations is to converge to these
vortex-filled equilibrium states (shared between the conservative and the
dissipative variant of the model), hence we use an unrealistically large value
of $\gamma$ to expedite this convergence.

Let us now summarize the main findings of this paper.

\begin{figure}[ptb]
\centering
\includegraphics[width=\textwidth]{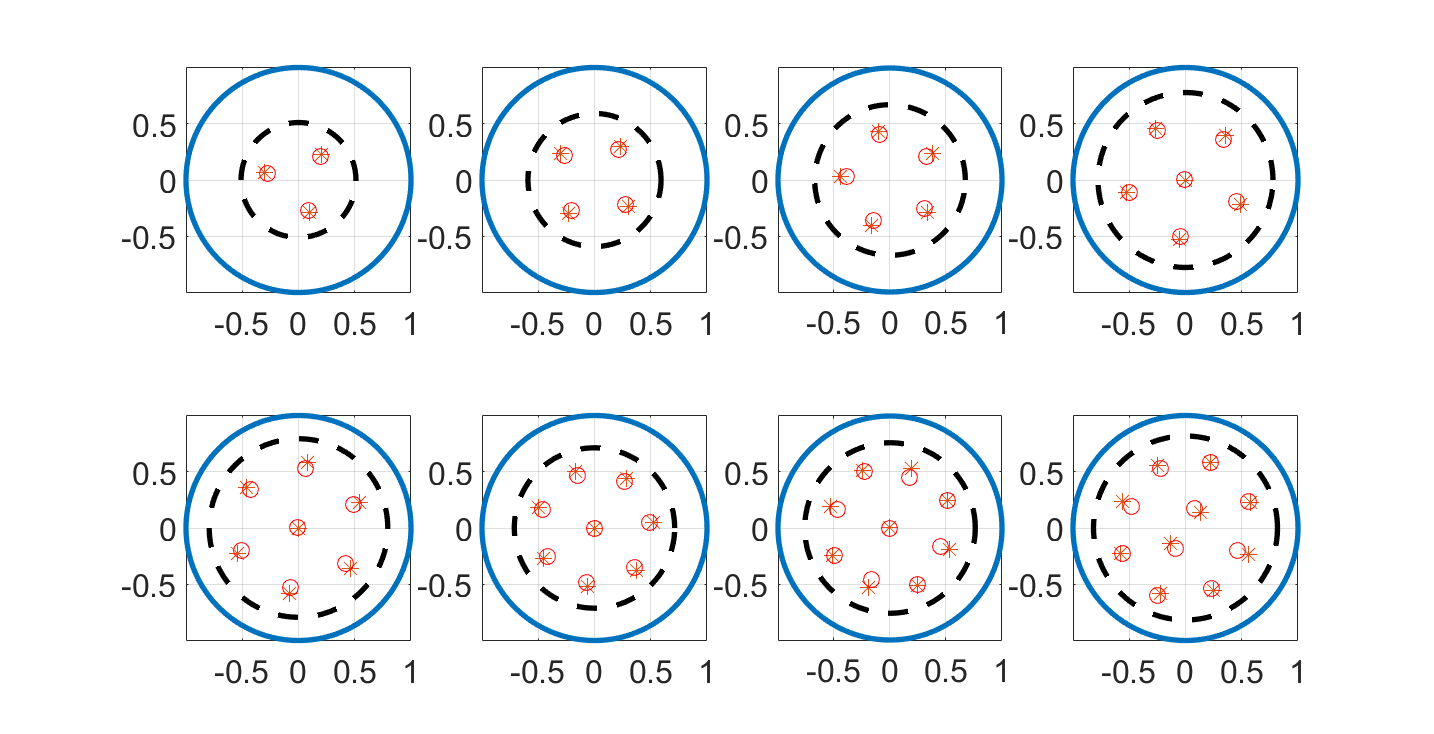}
\caption{Comparison of the steady state of PDE and ODE simulations. `$\ast$'
denotes the steady state of the ODE system (\ref{reduced}) whereas `o' is from
the PDE system (\ref{gp}). The parameters are chosen as: $\gamma
=1,~\kappa=0~,b=1,~\varepsilon=0.025$ and$~\Omega=29.51$ for 3 to 7 vortices
and $\Omega=36.89$ for 8 to 10 vortices. The dashed line represents the radius
prediction $a$ from (\ref{9:57}).}%
\label{fig:radial}%
\end{figure}

\begin{itemize}
\item[1.] \textbf{Reduced equations for vortex motion.} In \S \ref{S:2} and
\S \ref{S:3} we extend the asymptotic methods first developed in
\cite{weinan1994dynamics} to the case of a rotating trap. The presence of the
inhomogeneous trap introduces several complications, most notably the
inhomogeneous density background on top of which the vortices evolve (and
interact). The end result that we obtain through this analysis is the
following system for the motion of $N$ vortices whose positions are given by
$\xi_{j},\ j=1\ldots N:$
\begin{equation}
\gamma\log\left(  1/\varepsilon\right)  \xi_{jt}+\kappa\xi_{jt}^{\bot}=\left(
-\frac{2\Omega}{1+b^{2}}+\frac{2\log\left(  1/\varepsilon\right)  }{V(\xi
_{j})}\right)  \left(
\begin{array}
[c]{cc}%
1 & 0\\
0 & b^{2}%
\end{array}
\right)  \xi_{j}+2\sum_{k\neq j}\frac{(\xi_{j}-\xi_{k})}{|\xi_{j}-\xi_{k}%
|^{2}}\frac{V(\xi_{j})}{V(\xi_{k})}. \label{reduced}%
\end{equation}
Here and below, we use the notation $\left(  a,b\right)  ^{\perp}=(-b,a).$

We draw the reader's attention to the term $\frac{V(\xi_{j})}{V(\xi_{k})}$
which modifies the \textquotedblleft classical\textquotedblright%
\ Helmholtz-type vortex-to-vortex interaction of the form $\xi_{jt}^{\bot
}=\sum_{k\neq j}\frac{(\xi_{j}-\xi_{k})}{|\xi_{j}-\xi_{k}|^{2}}$. Equation
(\ref{reduced}) reduces to the \textquotedblleft classical\textquotedblright%
\ case (of Hamiltonian point vortex
motion) when $V=1,\gamma=0$ and $\Omega=0$, corresponding to a constant trap,
no rotation, and no damping. To our knowledge, this is the first time that
this additional term has been proposed and it incorporates in
a fundamental way the role of the potential (and also of the
anisotropy when the latter is present) towards screening
the inter-vortex interaction. In \cite{kolokolnikov2014tale}, the
same equation as (\ref{reduced})\ but without the term $\frac{V(\xi_{j}%
)}{V(\xi_{k})}$ was used to describe vortex dynamics in BEC. We show that our
modified equation (\ref{reduced})\ agrees with full numerical simulations of
the original GPE\ (\ref{GBP})\ much better, particularly in the case of
multiple vortices; relevant examples will be considered in
Figs.~\ref{fig:Nmax}, \ref{fig:radial}.

The remaining results in the paper follow from the analysis of the reduced
equation (\ref{reduced}).

\item[2.] \textbf{Large-}$N$ \textbf{vortex lattice density and radius for
isotropic potential.} Here, we extend the methods reported in
\cite{kolokolnikov2014tale} to derive the continuum limit density for the
steady state of (\ref{reduced}). In \S \ref{S:5} we show that in the large-$N$
limit, the radius $a$ of the vortex lattice is related to
$\Omega$, $N$, $\varepsilon$ via the formula
\begin{equation}
N\sim\frac{1}{\nu}\left(  \left(  -1-\frac{1}{2}\Omega\nu\right)  \ln
(1-a^{2})+2-2(1-a^{2})^{-1}\right)  ,\ \ \ N\gg1 \label{9:57}%
\end{equation}
where $\nu=1/\log\left(  1/\varepsilon\right)  .$ See Figure \ref{fig:radial},
where the asymptotic radius $a$ given by solving (\ref{9:57})\ is shown in
dashed curve, and a good agreement with full numerics is observed.

\item[3.] \textbf{Maximal admissible number of vortices}$.$ As we show in
\S \ref{S:5}, an immediate consequence of (\ref{9:57}) is the existence of a
fold-point bifurcation which results in the disappearence of some of the
vortices as $\Omega$ is decreased, as illustrated in figure \ref{fig:Nmax}.
Stated differently, for a fixed $\Omega,$ there is a maximum $N_{\max}$ such
that $N$-vortex lattice exists if and only if $N\leq N_{\max}$ where
\begin{equation}
N_{\max}=\frac{1}{\nu}\left\{  \left(  \Omega\nu+2\right)  \left(  \frac{1}%
{2}\ln(\Omega\nu+2)-\ln(2)-\frac{1}{2}\right)  +2\right\}  . \label{Nmax}%
\end{equation}
Figure \ref{fig:Nmax} illustrates this result.

\item[4.] \textbf{Stability of two vortices in the anisotropic case. }In
\S \ref{S:4} we study the stability of a two-vortex steady state with respect
to the above mentioned ODE dynamics. By symmetry, there are two equilibrium
states:\ the two vortices lying on major or minor axis. However, the
equilibrium along the minor axis is unstable~\cite{jan,goodman}.
Furthermore, a two vortex-state
on the major axis becomes unstable as $\Omega$ is decreased due to a fold
point bifurcation. We compute this bifurcation and compare this to numerics.
In paper \cite{aftalion2001vortices} a similar threshold was computed for the
anisotropic case from the energy point of view; this was also featured in the
work of~\cite{navarro2013dynamics} for the isotropic case, connecting the ODEs
with the GP PDE and also experimental results.

\item[5.] \textbf{High anisotropy, large }$N$\textbf{ limit (\S \ref{S:6}).
}Sufficiently high anisotropy \textquotedblleft pushes\textquotedblright\ all
the vortices to align along the major axis (see figure \ref{fig:anisotropy},
as well as~\cite{mcendoo2009small}; for some case examples with opposite
charges see~\cite{jan}). In the dual limit of high anisotropy and large $N,$
the steady state becomes essentially one-dimensional and we compute the
effective one-dimensional density using techniques involving the Chebychev
polynomials. As in the radially symmetric case,
the vortex \textquotedblleft
lattice\textquotedblright\ has a radius $a$ which, in the case $b\ll1,$ is
implicitly given via equation%
\begin{equation}
N\sim\frac{1}{\nu}\left(  \frac{\Omega\nu}{1+b^{2}}\frac{a^{2}}{2\sqrt
{1-a^{2}}}-\frac{(a^{2}-2)^{2}}{\nu(1-a^{2})^{\frac{3}{2}}}+1\right)  .
\label{12:10}%
\end{equation}

\item[6] \textbf{Maximal admissible number of vortices, high anisotropy
(\S \ref{S:6}). }Finally, as in the radially symmetric anisotropic case, we
compute $N_{\max,\text{1d}}$, the maximum number of vortices admissible for a
given $\Omega$ when the anisotropy is sufficiently high to align all vortices
along the major axis.\ It is obatined by maximizing (\ref{12:10}) which yields%
\begin{equation}
N_{\max,\text{1d}}=\frac{1}{\nu}\left(  1+3^{-3/2}\left(  \frac{\Omega\nu
}{1+b^{2}}-4\right)  \sqrt{1+2\frac{\Omega\nu}{1+b^{2}}}\right)  ,\ \ b\ll1
\label{12:08}%
\end{equation}

\end{itemize}

There have been two approaches to the dynamics of vortices in a trapped
condensate. The first approach relies on the fact that GP equation is the
Euler-Lagrange equation for the time-dependent Lagrangian functional under
variation of the wave function. If one is interested in an effective
description for the evolution of the vortex centers and how it varies upon
variation of one or more parameters, the resulting Lagrangian functional can
be used together with a multi-vortex ansatz to provide approximate Lagrangian
equations of motion~\cite{castin1999bose,kim2004dynamics,
aftalion2001vortices, kasamatsu2005vortices}. Another approach is to study GP
equation itself, which is the approach we take herein.
Due to the presence of two
length scales: the size of vortex core and the inter-vortex distance, it is
possible to employ the method of matched asymptotics \cite{weinan1994dynamics,
svidzinsky2000dynamics, svidzinsky2000stability, pismen1991motion,
rubinstein1994vortex}. This also leads to the derivation of dynamical
equations for the evolution of the vortex centers.

\section{Vortex dynamics}

\label{S:2}

We now derive vortex dynamics for (\ref{GBP}), following closely the
exposition of \cite{weinan1994dynamics}. We start by deriving the dynamics of
a single vortex, then expand our calculations to multiple vortices.

\subsection{Single vortex}

Suppose that the vortex center is located at $~\xi=(\zeta,\eta)\in%
\mathbb{R}
^{2}.$ Following \cite{weinan1994dynamics}, we decompose the solution into the
outer region $O(\varepsilon)$ away from the vortex center, and the inner
region near the vortex center. We will then use matched asymptotics to match
the two regions which will yield the equation of motion.


In the outer region, outside the vortex core $\left\vert x-\xi\right\vert \gg
O(\varepsilon)$, we decompose the solution into phase $\phi$ and amplitude
$u$:
\begin{equation}
w=ue^{i\phi}. \label{www1}%
\end{equation}
Substituting (\ref{www1}) into (\ref{GBP}) and separating the real and
imaginary part, we then obtain:
\begin{align}
\label{decomp}\gamma u_{t}+\kappa u\phi_{t}  &  =(\Delta u-u|\nabla\phi
|^{2})+\frac{1}{\varepsilon^{2}}\left(  V(x)-u^{2}\right)  u+\Omega
u\phi_{\theta}\\
-\kappa u_{t}+\gamma u\phi_{t}  &  =u\Delta\phi+2\nabla u\cdot\nabla
\phi-\Omega u_{\theta}.
\end{align}
We then expand $u$ and $\phi$ with respect to $\varepsilon$: $u=u_{0}+\varepsilon
u_{1}+\cdots$ and $\phi=\phi_{0}+\varepsilon\phi_{1}+\cdots$. The leading
order equations yield
\begin{subequations}
\label{leading}%
\begin{equation}
u_{0}=\sqrt{V(x)}%
\end{equation}
and
\begin{equation}
\gamma\phi_{0t}=\Delta\phi_{0}+\frac{1}{2}\frac{\nabla V}%
{V}\cdot\left(  2\nabla\phi_{0}-\Omega x^{\bot}\right)  . \label{phi_0}%
\end{equation}
 Assume that a vortex has charge +1, so that $\phi_{0}$ satisfies a
point boundary condition%
\end{subequations}
\begin{equation}
\phi_{0}\rightarrow\arg\left(  x-\xi\right)  \ \ \text{as }x\rightarrow\xi.
\label{phi0bc}%
\end{equation}
In order to match to the inner solution of the vortex inside the vortex core,
we need to understand in more detail the local behaviour of the outer solution
away from the vortex points. We first decompose $\phi_{0}$ as
\begin{equation}
\phi_{0}=S+\tilde{\phi}_{0}^{{}}%
\end{equation}
where $S$ is a regular solution (without any singularities)\ to
\begin{equation}
0=\Delta S+\frac{1}{2}\frac{\nabla V}{V}\cdot\left(  2\nabla S-\Omega x^{\bot
}\right)  .
\end{equation}
For the elliptic trap (\ref{elltrap}), its solution is given by
\begin{equation}
S(x)=\frac{\Omega}{2}\frac{b^{2}-1}{1+b^{2}}x_{1}x_{2}. \label{11:01}%
\end{equation}

Notice that this contribution vanishes in the isotropic limit of $b=1$. We
change to the moving coordinate $\tilde{x}=x-\xi(t)$ and denote by $(\tilde
{r},\tilde{\theta})$ the polar coordinates in moving coordinate. Then
(\ref{phi_0}) becomes:
\begin{equation}
\gamma\left(  \tilde{\phi}_{0t}-\xi_{t}\cdot\nabla\tilde{\phi}_{0}\right)
=\Delta\tilde{\phi}_{0}+\frac{\nabla V(\xi+\tilde{x})\cdot\nabla\tilde{\phi
}_{0}}{V(\xi+\tilde{x})}, \label{MovingCoordinate}%
\end{equation}
or, to leading order,
\begin{equation}
0\sim\Delta\tilde{\phi}_{0}+\left(  \gamma\xi_{t}^{\bot}+\frac{\nabla^{\bot
}V(\xi)}{V(\xi)}\right)  \cdot\nabla\tilde{\phi}_{0}. \label{10:53}%
\end{equation}
where we have assumed that the time-dynamics are sufficiently slow that
$\gamma\tilde{\phi}_{0t}$ can be discarded. In particular this is the case
near a stable equilibrium.

We now solve (\ref{10:53})\ iteratively near the singularity $\tilde
{x}\rightarrow0$. The leading-order solution must match the point-boundary
condition (\ref{phi0bc})\ which yields $\tilde{\phi}_{0}\sim\tilde{\theta}.$
Upon substituting $\tilde{\phi}_{0}\sim\tilde{\theta}+\phi_{01}$ we obtain%
\begin{equation}
0=\Delta\phi_{01}+\left(  \gamma\xi_{t}+\frac{\nabla V(\xi)}{V(\xi)}\right)
\cdot\left(  \frac{\tilde{x}^{\perp}}{\left\vert \tilde{x}\right\vert ^{2}%
}+\nabla\phi_{01}\right)  .
\end{equation}
The term $\nabla\phi_{01}$ is of smaller order than the other terms. Formal
expansion then yields%
\[
\phi_{01}=\frac{1}{2}(\log\tilde{r})\left(  \gamma\xi_{t}^{\bot}+\frac
{\nabla^{\bot}V(\xi)}{V(\xi)}\right)  \cdot\tilde{x}.
\]
Finally, at the next iteration we let $\tilde{\phi}_{0}\sim\tilde{\theta}%
+\phi_{01}+\phi_{02}.$ This yields $\phi_{02}\sim K\cdot\tilde{x}$ where the
vector $K$ depends on the vortex locations and will be determined later via
asymptotic matching. In summary, we obtain
\begin{equation}
\phi_{0}(\tilde{x},t)=S+\tilde{\theta}+\frac{1}{2}(\log\tilde{r})\left(  \gamma\xi_{t}^{\bot}+\frac{\nabla^{\bot}V(\xi)}{V(\xi
)}\right)  \cdot\tilde{x}+K\cdot\tilde{x}+\mathcal{O}(\tilde{r}^{2}\log
\tilde{r}) \label{11:03}%
\end{equation}
We now Taylor expand
the outer solution as $x\rightarrow\xi$. We have%
\begin{align*}
e^{i\phi_{0}}  &  =e^{i\left(  \tilde{\theta}+S(\xi)\right)  }\left(  \frac
{1}{2}(\log\tilde{r})\left(  \gamma\xi_{t}^{\bot}+\frac{\nabla^{\bot}V(\xi
)}{V(\xi)}\right)  \cdot\tilde{x}+K\cdot\tilde{x}\right)  +\mathcal{O}%
(\tilde{r}^{2}\log\tilde{r});\\
u_{0}  &  =\sqrt{V(\xi)}+\frac{\nabla V(\xi)\cdot\tilde{x}}{2\sqrt{V(\xi)}%
}+\mathcal{O}(\tilde{r}^{2});
\end{align*}
    
 This yields the following singularity behaviour for $w$ as $x\rightarrow\xi$:
\begin{equation}
\label{w_0Asy}w(\tilde{x},t)=e^{i\left(  \tilde{\theta}+S(\xi)\right)
}\left(  \sqrt{V(\xi)}+\frac{\nabla V(\xi)\cdot\tilde{x}}{2\sqrt{V(\xi)}%
}\right)  \left(  1+\frac{i}{2}(\log\tilde{r})\left(  \gamma\xi_{t}^{\bot
}+\frac{\nabla^{\bot}V(\xi)}{V(\xi)}\right)  \cdot\tilde{x}+i(K+\nabla
S)\cdot\tilde{x}\right)  +\mathcal{O}(\tilde{r}^{2}\log\tilde{r}%
)+\mathcal{O}(\varepsilon)
\end{equation}

Next we consider the inner region, let
\[
y=\displaystyle{\frac{x-\xi}{\varepsilon}}%
\]
and expand $w=W_{0}(y)+\varepsilon W_{1}(y)+\cdots$. In order to match each
order of $\varepsilon$, $W_{0}$, $W_{1}$ must satisfy:
\begin{align}
0  &  =\Delta_{y}W_{0}+V(\xi)W_{0}-|W_{0}|^{2}W_{0}\\
\left(  -(\gamma-\kappa i)\xi_{t}+i\Omega\xi^{\bot}\right)  \cdot\nabla
_{y}W_{0}-2\nabla V(\xi)\cdot yW_{0}  &  =\Delta_{y}W_{1}+V(\xi)W_{1}%
-|W_{0}|^{2}W_{1}-W_{0}\left(  W_{0}\overline{W_{1}}+W_{1}\overline{W_{0}%
}\right)
\end{align}
We scale out $V(\xi)$ by changing variables
\[
z=\sqrt{V(\xi)}y;\ \ W_{0}(y)=\sqrt{V(\xi)}U_{0}(z),~~W_{1}(y)=U_{1}(z)
\]
assuming that $\xi$ is slowly varying (so that it can be considered constant
along the scale of variation of $y$), in which case $U_{0},U_{1}$ satisfies:
\begin{align}
0  &  =\Delta_{z}U_{0}+U_{0}-|U_{0}|^{2}U_{0}\label{U_0}\\
\left(  -(\gamma-\kappa i)\xi_{t}+i\Omega\xi^{\bot}\right)  \cdot\nabla
_{z}U_{0}-\frac{\nabla V(\xi)\cdot z}{V(\xi)}U_{0}  &  =\Delta_{y}U_{1}%
+U_{1}-|U_{0}|^{2}U_{1}-U_{0}\left(  U_{0}\overline{U_{1}}+U_{1}%
\overline{U_{0}}\right)  . \label{U_1}%
\end{align}
We look for a vortex solution of $U_{0}$ in the form of $U_{0}(z)=f_{0}%
(R)e^{i\left(  \theta+S(\xi)\right)  },$ where $R,\theta$ denote the polar
coordinates of $z=Re^{i\theta}$. Then (\ref{U_0}) reduces to
\begin{equation}
f_{0}^{\prime\prime}+\frac{1}{R}f_{0}^{\prime}-\frac{1}{R^{2}}f_{0}%
+f_{0}(1-f_{0}^{2})=0 \label{f}%
\end{equation}
with the boundary condition:
\begin{equation}
f_{0}(0)=0,~~f_{0}(+\infty)=1. \label{fbc}%
\end{equation}
The solution to (\ref{f}, \ref{fbc})\ is well known to be unique
\cite{bethuel2012ginzburg}. The large $R$ expansion shows that $f_{0}$
satisfies
\begin{equation}
1-f_{0}^{2}-1/R^{2}=O(1/R^{4}),~~~R\rightarrow\infty.
\end{equation}
Let $U_{1}=f_{1}(R,\theta,t)e^{i\left(  \theta+S(\xi)\right)  }$. In terms of
$f_{1}$, (\ref{U_1}) becomes:
\begin{equation}
\left(  -(\gamma-\kappa i)\xi_{t}+i\Omega\xi^{\bot}\right)  \cdot\left(
f_{0}^{\prime}\nabla_{z}R+if_{0}\nabla_{z}\theta\right)  -\frac{\nabla
V(\xi)\cdot z}{V(\xi)}f_{0}=\Delta_{z}f_{1}+2i\left(  \nabla_{z}f_{1}%
\cdot\nabla_{z}\theta\right)  -\frac{1}{R^{2}}f_{1}+f_{1}(1-2f_{0}^{2}%
)-f_{0}^{2}\overline{f_{1}}%
\end{equation}
We then decompose further $f_{1}=A(R)\cos\theta+B(R)\sin\theta$ and separate
real and imaginary parts:
\[
A=A_{r}+iA_{i},~~~B=B_{r}+iB_{i}%
\]
to obtain the following equations for $A_{r},A_{i},B_{r},B_{i}$:
\begin{align}
-\frac{V_{x_{1}}(\xi)R}{V(\xi)}f_{0}-\gamma\zeta_{t}f_{0}^{\prime}%
-\frac{\Omega\zeta+\kappa\eta_{t}}{R}f_{0}  &  =A_{r}^{\prime\prime}+\frac
{1}{R}A_{r}^{\prime}+(1-3f_{0}^{2}-\frac{2}{R^{2}})A_{r}-\frac{2B_{i}}{R^{2}%
}\\
-\frac{V_{x_{2}}(\xi)R}{V(\xi)}f_{0}-\gamma\eta_{t}f_{0}^{\prime}-\frac
{\Omega\eta-\kappa\zeta_{t}}{R}f_{0}  &  =B_{r}^{\prime\prime}+\frac{1}%
{R}B_{r}^{\prime}+(1-3f_{0}^{2}-\frac{2}{R^{2}})B_{r}+\frac{2A_{i}}{R^{2}}\\
\frac{-\gamma\eta_{t}f_{0}}{R}-\Omega\eta f_{0}^{\prime}+\kappa\zeta_{t}%
f_{0}^{\prime}  &  =A_{i}^{\prime\prime}+\frac{1}{R}A_{i}+(1-f_{0}^{2}%
-\frac{2}{R^{2}})A_{i}+\frac{2B_{r}}{R^{2}}\\
\frac{\gamma\zeta_{t}f_{0}}{R}+\Omega\zeta f_{0}^{\prime}+\kappa\eta_{t}%
f_{0}^{\prime}  &  =B_{i}^{\prime\prime}+\frac{1}{R}B_{i}+(1-f_{0}^{2}%
-\frac{2}{R^{2}})B_{i}+\frac{2A_{r}}{R^{2}}.
\end{align}
We are concerned about the behaviour of the solutions of these equations at
infinity. As $R\rightarrow\infty$, we have:
\begin{align}
-\frac{V_{x_{1}}(\xi)R}{V(\xi)}\left(  1-\frac{1}{R^{2}}\right)  -\frac
{\Omega\zeta+\kappa\eta_{t}}{R}  &  =A_{r}^{\prime\prime}+\frac{1}{R}%
A_{r}^{\prime}+(-2+\frac{1}{R^{2}})A_{r}-\frac{2B_{i}}{R^{2}}+\mathcal{O}%
\left(  \frac{1}{R^{3}}\right) \\
-\frac{V_{x_{2}}(\xi)R}{V(\xi)}\left(  1-\frac{1}{R^{2}}\right)  -\frac
{\Omega\eta-\kappa\zeta_{t}}{R}  &  =B_{r}^{\prime\prime}+\frac{1}{R}%
B_{r}^{\prime}+(-2+\frac{1}{R^{2}})B_{r}+\frac{2A_{i}}{R^{2}}+\mathcal{O}%
\left(  \frac{1}{R^{3}}\right) \\
\frac{-\gamma\eta_{t}}{R}  &  =A_{i}^{\prime\prime}+\frac{1}{R}A_{i}^{\prime
}-\frac{1}{R^{2}}A_{i}+\frac{2B_{r}}{R^{2}}+\mathcal{O}\left(  \frac{1}{R^{3}%
}\right) \\
\frac{\gamma\zeta_{t}}{R}  &  =B_{i}^{\prime\prime}+\frac{1}{R}B_{i}^{\prime
}-\frac{1}{R^{2}}B_{i}-\frac{2A_{r}}{R^{2}}+\mathcal{O}\left(  \frac{1}{R^{3}%
}\right)  .
\end{align}
By expressing the solutions in a power series of $R$ and $\log R$ for large
$R,$ we obtain\bes%
\begin{equation}
A_{r}=\frac{V_{x_{1}}(\xi)R}{2V(\xi)}-\left(  \frac{\gamma\zeta_{t}}{2}%
+\frac{V_{x_{1}}(\xi)}{2V(\xi)}\right)  \frac{\log R}{R}+\mathcal{O}(\frac
{1}{R})
\end{equation}%
\begin{equation}
B_{r}=\frac{V_{x_{2}}(\xi)R}{2V(\xi)}+\left(  -\frac{\gamma\eta_{t}}{2}%
-\frac{V_{x_{2}}(\xi)}{2V(\xi)}\right)  \frac{\log R}{R}+\mathcal{O}(\frac
{1}{R})
\end{equation}%
\begin{equation}
A_{i}=\left(  -\frac{\gamma\eta_{t}}{2}-\frac{V_{x_{2}}(\xi)}{2V(\xi)}\right)
R\log R-\frac{1}{2}\Omega\eta R+\frac{\kappa\zeta_{t}R}{2}+\mathcal{O}(\log R)
\end{equation}%
\begin{equation}
B_{i}=\left(  \frac{\gamma\zeta_{t}}{2}+\frac{V_{x_{1}}(\xi)}{2V(\xi)}\right)
R\log R+\frac{1}{2}\Omega\zeta R+\frac{\kappa\eta_{t}R}{2}+\mathcal{O}(\log R)
\end{equation}
Putting these together, we get for $R\gg1,$ \ees%
\begin{equation}
U_{1}(z,t)=e^{i\theta+S(\xi)}\left[  \frac{\nabla V(\xi)z}{2V(\xi)}+\frac
{i}{2}(\log R)\left(  \gamma\xi_{t}^{\bot}+\frac{\nabla^{\bot}V(\xi)}{V(\xi
)}\right)  \cdot z+\frac{i}{2}\Omega\xi^{\bot}\cdot z+\frac{\kappa\xi_{t}}%
{2}\cdot z\right]  .
\end{equation}
Therefore, as $R\rightarrow\infty$, the asymptotic behaviour of the inner
solution is given by:
\begin{equation}
W_{0}+\varepsilon W_{1}=e^{i\theta+iS(\xi)}\left(  \sqrt{V(\xi)}%
f_{0}(R)+\varepsilon\left[  \frac{\nabla V(\xi)z}{2V(\xi)}+\frac{i}{2}(\log
R)\left(  \gamma\xi_{t}^{\bot}+\frac{\nabla^{\bot}V(\xi)}{V(\xi)}\right)
\cdot z+\frac{i}{2}\Omega\xi^{\bot}\cdot z+\frac{\kappa\xi_{t}}{2}\cdot
z\right]  \right).  \label{w_inner}%
\end{equation}
To match (\ref{w_inner}) with (\ref{w_0Asy}), we recall that $\tilde{x}%
=\frac{\varepsilon z}{\sqrt{V(\xi)}},~\tilde{r}=\frac{\varepsilon R}%
{\sqrt{V(\xi)}}$. Asymptotic matching then yields%
\[
\frac{i}{2}\Omega\xi^{\bot}+\frac{\kappa\xi_{t}}{2}\sim i(K+\nabla S)+\frac
{i}{2}(\log\frac{\varepsilon}{\sqrt{V}})\left(  \gamma\xi_{t}^{\bot}%
+\frac{\nabla^{\bot}V(\xi)}{V(\xi)}\right)
\]
or
\begin{equation}
\gamma\xi_{t}^{\bot}-\kappa\nu\xi_{t}\sim\nu\left(  -\Omega\xi^{\bot}+2\nabla
S+2K\right)  -\frac{\nabla^{\bot}V(\xi)}{V(\xi)}. \label{409}%
\end{equation}
where ${\nu=\frac{1}{\log\left(  \sqrt{V(\xi)}/(\varepsilon) \right)  }%
\sim\frac{1}{\log\left(  1/\varepsilon\right)  }.}$ The quantity $K$ will be
determined in \S \ref{S:3} below through asymptotic matching, and incorporates
multi-vortex interactions. In the case of a single vortex, we will show that
$K$ is bounded and thus asymptotically small compared to the other terms. In
addition we recall from (\ref{11:01}, \ref{elltrap})\ that $\nabla
S=\frac{\Omega}{2}\frac{b^{2}-1}{1+b^{2}}\left(  \eta,\zeta\right)  $ and
$\frac{\nabla^{\bot}V(\xi)}{V(\xi)}=\frac{2\left(  b^{2}\eta,-\zeta\right)
}{1-\zeta^{2}-b^{2}\eta^{2}}$ so that (\ref{409})\ simplifies to%
\begin{equation}
\gamma\xi_{t}^{\bot}-\kappa\nu\xi_{t}=\left(  \frac{-2\Omega\nu}{1+b^{2}%
}+\frac{2}{1-\zeta^{2}-b^{2}\eta^{2}}\right)  \left(  -b^{2}\eta,\zeta\right)
\end{equation}
or equivalently,%
\begin{equation}
\gamma\xi_{t}+\kappa\nu\xi_{t}^{\bot}=\left(  \frac{-2\Omega\nu}{1+b^{2}%
}+\frac{2}{1-\zeta^{2}-b^{2}\eta^{2}}\right)  \left(
\begin{array}
[c]{cc}%
1 & 0\\
0 & b^{2}%
\end{array}
\right)  \xi\label{no_bot}%
\end{equation}
An immediate corrollary of (\ref{no_bot}) is that a single vortex at the
center $\xi=0$ is stable if and only if $\Omega>\Omega_{1}$ where%
\begin{equation}
\Omega_{1}=\frac{1+b^{2}}{\nu}. \label{Omega1}%
\end{equation}
As a consequence, no stable votices exist below the critical rotation rate
$\Omega<\Omega_{1}.$ The exact same critical rate was previously derived in
\cite{aftalion2001vortices} using energy methods, as well as, e.g., discussed
in~\cite{dep_amrx} in the context of bifurcation theory. Asymptotically, this
agrees with the numerical simulations of the full PDE\ system (\ref{gp});
however, there are nontrivial corrections on this frequency that were
addressed, e.g., in the work of~\cite{middelkamp}.

\subsection{Multiple vo\textbf{rtices}}

\label{S:3} We now look for approximate solution of (\ref{GBP}) with $N$
vortices in the location $\xi_{j},~j=1..N$, where all of the vortices bear the
same charge $+1.$ Proceeding in the same way as for a single vortex, we attempt to
study the dynamics of $N$ such vortices. The inner solution $W_{0}$ near the
core of vortices is the same as for a single vortex. In the outer region,
$\tilde{\phi}_{0}$ still satisfies the equation (\ref{phi_0}) but with $N$
point boundary conditions $\tilde{\phi}_{0}\sim\arg\left(  x-\xi_{j}\right)  $
as $x\rightarrow\xi_{j}.$ The singularity analysis of the outer region near
$\xi_{j}$ is identical to the derivation of (\ref{11:03})\ with the end
result
\begin{equation}
\tilde{\phi}_{0}(\tilde{x},t)\sim\tilde{\theta}+\frac{1}{2}(\log\tilde
{r})\left(  \gamma\xi_{t}^{\bot}+\frac{\nabla^{\bot}V(\xi_{j})}{V(\xi_{j}%
)}\right)  \cdot\tilde{x}+K_{j}\cdot\tilde{x} \label{320}%
\end{equation}
where $\tilde{x}=x-\xi_{j},\ \tilde{r}=\left\vert \tilde{x}\right\vert $ with
$\tilde{x}\rightarrow0.$ The multi-vortex analogue for (\ref{no_bot})\ is
\begin{equation}
\gamma\xi_{jt}+\kappa\nu\xi_{jt}^{\bot}=\left(  \frac{-2\Omega\nu}{1+b^{2}%
}+\frac{2}{1-\xi_{j1}^{2}-b^{2}\xi_{j2}^{2}}\right)  \left(
\begin{array}
[c]{cc}%
1 & 0\\
0 & b^{2}%
\end{array}
\right)  \xi_{j}-\nu2K_{j}^{\perp}.
\end{equation}
It remains to determine the constants $K_{j}$ via asymptotic matching. In the
outer region, $\tilde{\phi}_{0}$ satisfies $0\sim\Delta\tilde{\phi}_{0}%
+\frac{\nabla V(x)\cdot\nabla\tilde{\phi}_{0}}{V(x)}$ or
equivalently,\bes\label{1055}%
\begin{equation}
\nabla\cdot\left(  V(x)\nabla\tilde{\phi}_{0}\right)  =0, \label{1053}%
\end{equation}
with $N$ point-boundary conditions
\begin{equation}
\tilde{\phi}_{0}\sim\arg\left(  x-\xi_{j}\right)  \text{ \ as \ }%
x\rightarrow\xi_{j},\ \ \ \ j=1\ldots N \label{1054}%
\end{equation}
\ees

In the derivation that follows, we will assume that the vortices are close to
each other, separated by a small distance of $O(1/\log(1/\varepsilon))$.
Similar to a computation in \cite{sheehy2004vortices}, the leading-order
solution to (\ref{1055}) is then given by\footnote{The full solution to
(\ref{1055})\ is $\nabla\tilde{\phi}=\sum_{k}\frac{V(\xi_{k})}{V(x)}\nabla
\arg\left(  x-\xi_{k}\right)  +\frac{\nabla^{\perp}\psi}{V(x)}$  where $\psi$ is
chosen in such a way as to satisfy the solvability condition to make
$\tilde{\phi}$ a true gradient. In particular, $\frac{\nabla^{\perp}\psi
}{V(x)}$ is zero when $V$ is constant. More generally, $\psi$ satisfies
$\nabla\cdot\left(  \frac{\nabla\psi}{V(x)}\right)  =\sum_{k}\nabla\left(
\frac{V(\xi_{k})}{V(x)}\right)  \cdot\nabla^{\perp}\arg\left(  x-\xi
_{k}\right)  $. In what follows, we assume that the vortices are close to each
other in which case the term $\nabla\arg\left(  x-\xi_{k}\right)  $ dominates
and $\psi$ provides a higher-order contribution which we can ignore.}
\[
\nabla\tilde{\phi}\sim\sum_{k}\frac{V(\xi_{k})}{V(x)}\nabla\arg\left(
x-\xi_{k}\right)  .
\]
Letting $x\rightarrow\xi_{j}$ we then obtain%
\[
\nabla\tilde{\phi}\sim\nabla\tilde{\theta}+\sum_{k\neq j}\frac{V(\xi_{k}%
)}{V(\xi_{j})}\nabla\arg\left(  \xi_{k}-\xi_{j}\right)  .
\]
Matching with (\ref{320}) then yields%
\[
K_{j}=\sum_{k\neq j}\frac{V(\xi_{k})}{V(\xi_{j})}\nabla\arg\left(  \xi_{k}%
-\xi_{j}\right)  =-\sum_{k\neq j}\frac{V(\xi_{k})}{V(\xi_{j})}\frac{(\xi
_{j}-\xi_{k})^{\perp}}{|\xi_{j}-\xi_{k}|^{2}}.
\]

This yields the final result, which we summarize as follows:%

\begin{equation}
\gamma\xi_{jt}+\nu\kappa\xi_{jt}^{\bot}\sim\left(  -\frac{2\nu\Omega}{1+b^{2}%
}+\frac{2}{1-\xi_{j1}^{2}-b^{2}\xi_{j2}^{2}}\right)  \left(
\begin{array}
[c]{cc}%
1 & 0\\
0 & b^{2}%
\end{array}
\right)  \xi_{j}+2\sum_{k\neq j}\frac{\nu(\xi_{j}-\xi_{k})}{|\xi_{j}-\xi
_{k}|^{2}}\frac{V(\xi_{j})}{V(\xi_{k})}. \label{n_vortices}%
\end{equation}
This concludes the derivation of formula (\ref{reduced}), which is the
starting point for all the subsequent results of this paper. The fundamental
element of novelty in our dynamical equations lies in the treatment of the
interaction terms, as both the anisotropic and the dissipative cases have been
recently considered in a similar vein from the viewpoint of effective particle
dynamics; see, e.g.,~\cite{goodman} and~\cite{prouk} for respective examples.
In what follows, we will proceed to analyze the resulting systems for $N=2$,
as well as for general $N$ number of vortices for both isotropic and
anisotropic traps, comparing the conclusions to those stemming from direct
numerical simulations.

\section{Multi-vortex lattice density, isotropic trap.}

\label{S:5}

We start by considering isotropic parabolic potential ($b=1$) in the regime
where the number of vortices $N$ is large. As demonstrated in experiments
\cite{abo2001observation, raman2001vortex}, in this case the vortices settle
to a hexagonal \textquotedblleft crystal lattice\textquotedblright%
\ configurations such as shown in Figure \ref{fig:Nmax}. Our goal is to
estimate the asymptotic density of the resulting lattice using techniques
similar to those of \cite{kolokolnikov2014tale}. As a direct consequence, this
computation will also yield the maximum allowed number $N_{\max}$ of vortices
as a function of system parameters.

We start with the ODE system (\ref{reduced}) that describes the evolution of
multiple vortex centers. Since we are interested in the fundamental (stable
equilibrium) states $\xi_{j}\left(  t\right)  \rightarrow\xi_{j},$ we only
consider the overdamped regime (i.e. imaginary time integration)
$\gamma\rightarrow\infty.$ Equivalently, by rescaling the time, in the case of
the isotropic potential $\left(  b=1\right)  $ the system (\ref{n_vortices}%
)\ may be written as%

\begin{equation}
\xi_{j\tau}=\left(  -\nu\Omega+\frac{2}{1-|\xi_{j}|^{2}}\right)  \xi_{j}%
+2\nu(1-|\xi_{j}|^{2})\sum_{k\neq j}\frac{\xi_{k}-\xi_{j}}{|\xi_{k}-\xi
_{j}|^{2}}\frac{1}{1-|\xi_{k}|^{2}}. \label{n_vortices3}%
\end{equation}
Being interested in the limit of large $N$ (in which case a near ``continuum
of vortices'' emerges) and following \cite{kolokolnikov2014tale}, we
coarse-grain the system. This is done by defining a particle density according
to:
\begin{equation}
\rho(x)=\sum\delta(x-\xi_{k}). \label{10:52}%
\end{equation}
Equation (\ref{n_vortices3}) can then be written as $\xi_{j\tau}=v(\xi_{j})$
where the velocity $v$ is given by\bes\label{3:29}%
\begin{equation}
v(x)=\left(  -\nu\Omega+\frac{2}{1-|x|^{2}}\right)  x+2\nu(1-|x|^{2}%
)\int_{R^{2}}\frac{x-y}{|x-y|^{2}}\frac{1}{1-|y|^{2}}\rho(y)dy.
\label{transport}%
\end{equation}
In the continuum limit $N\rightarrow\infty,$ this equation is coupled to the
conservation of mass,%
\begin{equation}
\rho_{\tau}(x,\tau)+\nabla_{x}\cdot\left(  v(x)\rho(x,\tau)\right)  =0
\end{equation}
\ees Together, (\ref{3:29}) describe the vortex density evolution in the limit
$N\rightarrow\infty$ for the overdamped regime (\ref{n_vortices3}). Similarly
to the analysis of~\cite{kolokolnikov2014tale}, it can be found that the
resulting steady state density $\rho$ is compactly supported. Assuming that
the density is radial, it is possible to compute the steady state $\rho\left(
x,t\right)  =\rho\left(  \left\vert x\right\vert \right)  $ and its radial
support explicitly using techniques from \cite{kolokolnikov2014tale}, as we
now show. Assume that the density is supported on a disk of radius $a,$ so
that $\rho(r)=0$ for $r>a$ and $\rho(r)>0$ for $0\leq r<a.$ A key identity is
\begin{equation}
\int_{R^{2}}\frac{x-y}{|x-y|^{2}}g(\left\vert y\right\vert )dy=x\frac{2\pi
}{r^{2}}\int_{0}^{r}g(s)sds, \label{gid}%
\end{equation}
which holds for any integrable function $g(r)$. \begin{figure}[tb]
\caption{ Number of vortices $N$ as a function the vortex lattice radius $a$.
Note the appearence of a maximum $N_{\max}$ corresponding to the maximum
admissible number of vortices.}%
\label{fig:Na}%
\centering\includegraphics[width=0.5\textwidth]{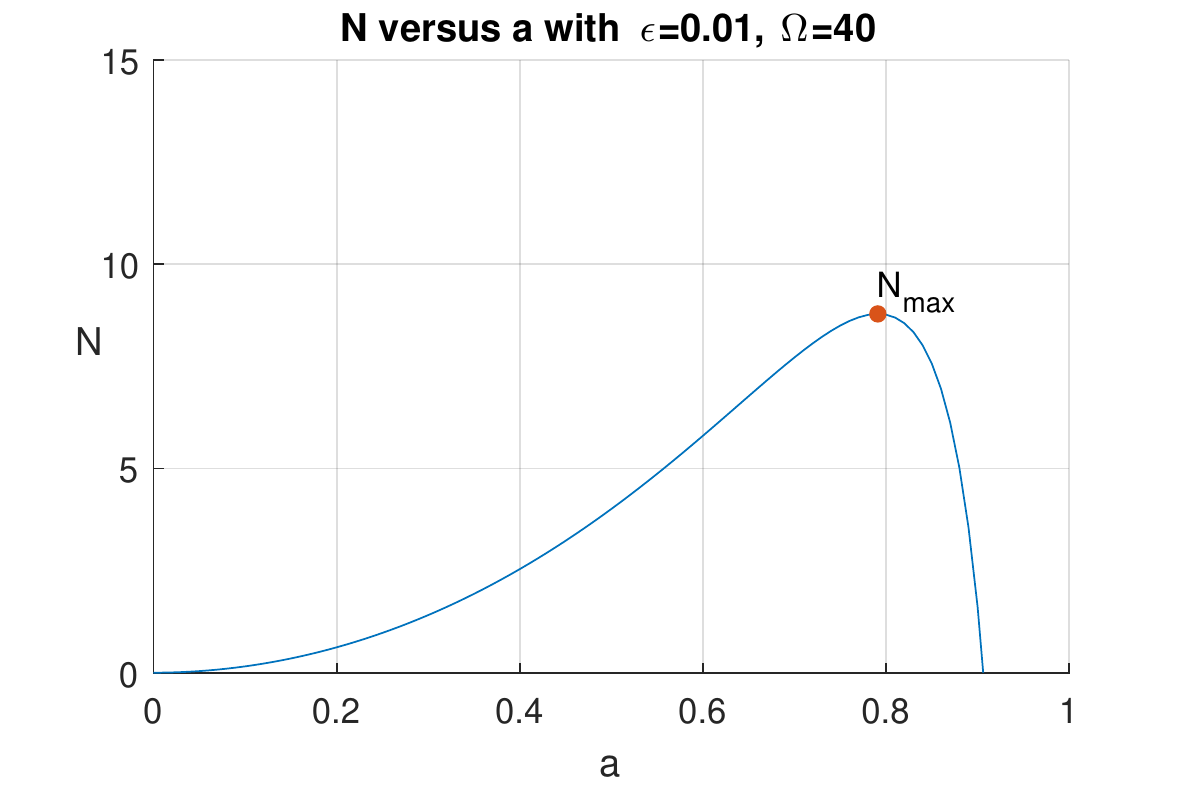}
\end{figure}

Applying (\ref{gid}) to (\ref{transport}) then yields
\begin{equation}
v(x)=\left(  -\nu\Omega+\frac{2}{1-r^{2}}+\frac{4\pi\nu(1-r^{2})}{r^{2}}%
\int_{0}^{r}\frac{1}{1-s^{2}}\rho(s)sds\right)  x.
\end{equation}
Inside the support $r<a$, we set $v=0.$ Upon differentiating with respect to
$r$ we obtain%
\begin{equation}
\rho(r)=\frac{1}{4\pi\nu}\left(  -\frac{2\Omega\nu r}{(1-r^{2})}-\frac
{4}{1-r^{2}}+\frac{8}{(1-r^{2})^{2}}\right)  . \label{rho_iso}%
\end{equation}
Note from (\ref{10:52})\ that the total mass is $N$. Since we assumed that the
density is supported on $\left\vert x\right\vert <a,$ this leads to an
additional constraint%
\begin{equation}
2\pi\int_{0}^{a}\rho(s)sds=N. \label{Constrain_iso}%
\end{equation}
Combining (\ref{rho_iso}) and (\ref{Constrain_iso}), we obtain an explicit
relationship between the support radius $a$ and $N$, which is given by
Eq.~(\ref{9:57}).
%

A typical graph of $N$ versus $a$ is shown in Figure \ref{fig:Na}. Note that
this graph attains the maximum which we compute by setting $\partial
N/\partial a=0.$ This maximum $N_{\max}$ is attained at $a=\frac{\sqrt
{\Omega\nu-2}}{\sqrt{\Omega\nu+2}}$ and has an explicit expression given by
Eq.~(\ref{Nmax}).

Formula (\ref{Nmax}) is one of the main results of this paper:\ it gives the
maximum admissible number of vortices for a given rotation rate $\Omega$.
Figure \ref{fig:Nmax} compares this formula (see solid curve in figure
\ref{fig:Nmax}(b)) with both the full PDE\ simulations as well as the
simulation of ODEs (\ref{reduced}), from which this formula is derived.

To generate the curve \textquotedblleft ODE\textquotedblright, we simulated
the ODE system (\ref{reduced}), starting with $\Omega=125$ and $N=80.$ A
simple forward
Euler method was found to be sufficient and was used with the stepsize $dt=0.01.$ We very gradually decreased
$\Omega$ until such time that one of the particles escaped the trap (i.e.
$\left\vert x_{j}\right\vert \geq1$ for some $j$). When this occured, we
decreased $N$ by one, and recorded the corresponding $\Omega.$ The points
where $N$ drops corresponds to the \textquotedblleft
disappearence\textquotedblright\ of vortices, and are indicated by step
discontinuties of the curve \textquotedblleft ODE\textquotedblright\ in the
figure. For the PDE, we simulated (\ref{gp})\ using FlexPDE inside a disk of
radius $1.3\ $with Dirichlet boundary conditions: $w(x)=0$ when $\left\vert
x\right\vert =1.3$. Since the solution decays rapidly outside the trap
$\left\vert x\right\vert >1$, this radius was sufficient to discard any
boundary effects (we also validated that by increasing the domain radius
and ensuring that that did not
affect the solution). We used the winding number of $w$ around the contour
$\left\vert x\right\vert =R,$ where $R$ is the radius chosen in such a way
that $\left\vert u\right\vert <10^{-4}$ for all $x\geq R$, to compute the
number of vortices for any given snapshot. The steps in the graph correspond
to values of $\Omega$ where the winding number is decreased. For both PDE and
ODE computations, we made sure that $\Omega$ was decreasing much slower than
any transient dynamics, so that the system is in a quasi steady state, except
at the points where the vortices \textquotedblleft disappear\textquotedblright.

We remark that the expression (\ref{Nmax}) for $N_{\max}$ is an
\emph{asymptotic }result, in the dual limit $\varepsilon\rightarrow0$ and
$N\gg1.$ In other words, it is an approximation to the true upper bound and
should not be considered as an upper bound itself. In particular the crossings
of \textquotedblleft continuum\textquotedblright\ and \textquotedblleft
PDE\textquotedblright\ curves in Figure \ref{fig:Nmax}(b) does not contradict
our results:\ we only claim that the curves \textquotedblleft
continuum\textquotedblright\ and \textquotedblleft PDE\textquotedblright%
\ asymptote to each other for large $N$. On the other hand, there is a limit
to the validity of the asymptotic results:\ if there are too many vortices,
their inter-vortex
distance decreases and asymptotics eventually start to fail. In
practice, this imposes a restriction of how big $\Omega$ can be until the
asymptotics start to fail.

It should be noted here as regards $\nu$ that its logarithmic factor involves
$\log(1/\varepsilon)$, while in connection with the numerical
work~\cite{middelkamp}, a more accurate factor of $\log(A/\varepsilon)$ has
been proposed, yielding improved agreement with the precession frequency.

It is worthwhile to also mention
that Aftalion and Du \cite{aftalion2001vortices} derived a different
formula for the threshold $N_{\max}$ but using the variational framework; see
formula (3.4)\ in \cite{aftalion2001vortices}. In our notation, this formula
can be rewritten as%
\begin{equation}
N_{\max,\text{Aftalion/Du}}=1+\left(  \Omega-\frac{2}{\nu}\right)  \frac
{1}{\log\left(  2/\nu\right)  }. \label{NmaxAD}%
\end{equation}
It is also shown in Figure \ref{fig:Nmax}. Unlike our formula
(\ref{NmaxAD}) is linear in $\Omega$ and while reasonably accurate for a small
number of vortices, it becomes progressively less acurate for large $\Omega$.

Finally, let us mention that a similar computation was done in
\cite{kolokolnikov2014tale} for a simplified version of the vortex equations
of motion that did not incorporate the trap density in vortex-to-vortex
interactions suggested in \cite{navarro2013dynamics}, namely
\begin{equation}
z_{j\tau}=\left(  -\nu\Omega+\frac{2}{1-|z_{j}|^{2}}\right)  z_{j}+2\nu
\sum_{k\neq j}\frac{z_{k}-z_{j}}{|z_{k}-z_{j}|^{2}}.
\end{equation}
For this simplified system, a similar analysis (see
\cite{kolokolnikov2014tale}, section 4)\ yields the formula
\begin{equation}
N_{\max\text{,CKK}}=\frac{1}{\nu}\left(  \sqrt{\frac{\Omega\nu}{2}}-1\right)
^{2}. \label{Nmax2}%
\end{equation}
In fact, formulas (\ref{Nmax})\ and (\ref{Nmax2})\ both agree near $\Omega
\nu=2$ as can be seen by expanding\ in Taylor series around $\Omega\nu=2;$ in
this regime, $\nu N_{\max}$ is small, the radius $a$ is also small and both
formulas yield $\nu N_{\max}=\frac{1}{16}\left(  \Omega\nu-2\right)
^{2}+O(\left(  \Omega\nu-2\right)  ^{3})$ with $a\sim\sqrt{\Omega\nu
-2}+o(\sqrt{\Omega\nu-2}).$ However the two deviate significantly for larger
values of $N.$

\section{Two vortices, anisotropic trap}

\label{S:4}

Let us now investigate in some more detail the case of two vortices in an
anisotropic trap $\left(  b\neq1\right)  $. In the isotropic case $\left(
b=1\right)  $, a basic steady state configuration consists of two antipodal
vortices along \emph{any } line through the center due to the
rotational invariance of the model. It should be highlighted,
however, that the work of~\cite{navarro2013dynamics,zampetaki} revealed that
this configuration is only stable within a range of distances of the antipodal
pair from the origin. Beyond a critical threshold, the energetically favored
state becomes an asymmetric one. On the other hand, even for the antipodal
states, the introduction of the anisotropy breaks the rotational symmetry,
leading to two possible steady states:\ either vortex centers lie on the
x-axis or on the y-axis. Both configurations may be admissible as steady
states. However the stability analysis below will show that only  the
configuration with two vortices along the longest axis of the ellipse
$x^{2}+by^{2}=1$ is stable, the other configuration being unstable. This is in
line with earlier works in the case of oppositely charged vortices; see,
e.g.,~\cite{jan}.

First, consider two votices in a stable configuration along the $x$-axis, with
coordinates $\xi_{1}=\left(  r,0\right)  $ and $\xi_{2}=\left(  -r,0\right)
.$ Upon substituting into the equation of motion (\ref{reduced}) we obtain an
algebraic equation for $r,$%
\begin{equation}
\left(  -\frac{\nu\Omega}{1+b^{2}}+\frac{1}{1-r^{2}}\right)  r+\frac{\nu}%
{2r}=0. \label{c_0}%
\end{equation}
This equation is quadratic in $r^{2}$, and admits two positive solutions
$r_{\pm}$ with $r_{-}<r_{+},$ provided that $\Omega>\Omega_{2}$ where%
\begin{equation}
\Omega_{2}=\frac{1}{\nu}\frac{1+b^{2}}{2}\left(  \sqrt{2}+\sqrt{\nu}\right)
^{2}. \label{Omega2}%
\end{equation}
There is a fold point at $\Omega=\Omega_{2}$ and the solution disappears when
$\Omega<\Omega_{2}$. This was already observed in the work
of~\cite{navarro2013dynamics} in the case of an isotropic trap. Note that to
leading order in $\nu$, $\Omega_{2}\sim\left(  1+b^{2}\right)  /\nu,$ which
agrees with the stability threshold for a single spike $\Omega_{1}$, see
(\ref{Omega1}). We also remark that the same formula for $\Omega_{2}$ holds
for two vortices along the $y-$axis. This can be seen as follows:\ assume that
the equilibrium is at $\left(  0,\pm\hat{r}\right)  .$ By rescaling $\hat
{r}=r/b,$ we find that $r$ then satisfies (\ref{c_0}), so that the fold point
$\Omega_{2}$ is the same whether the vortices are along $x-$ or $y-$ axis.

\begin{figure}[tb]
\centering\begin{subfigure}[b]{0.4\textwidth}
\includegraphics[width=\textwidth]{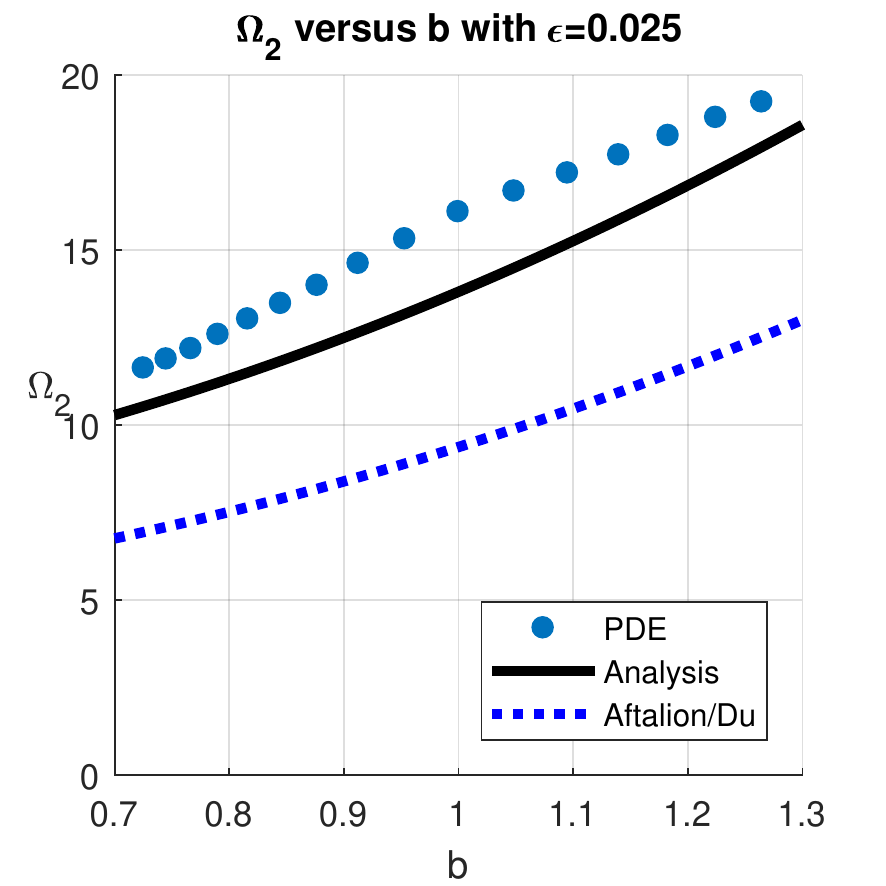}
\caption{}
\label{fig:t1}
\end{subfigure}\begin{subfigure}[b]{0.4\textwidth}
\includegraphics[width=\textwidth]{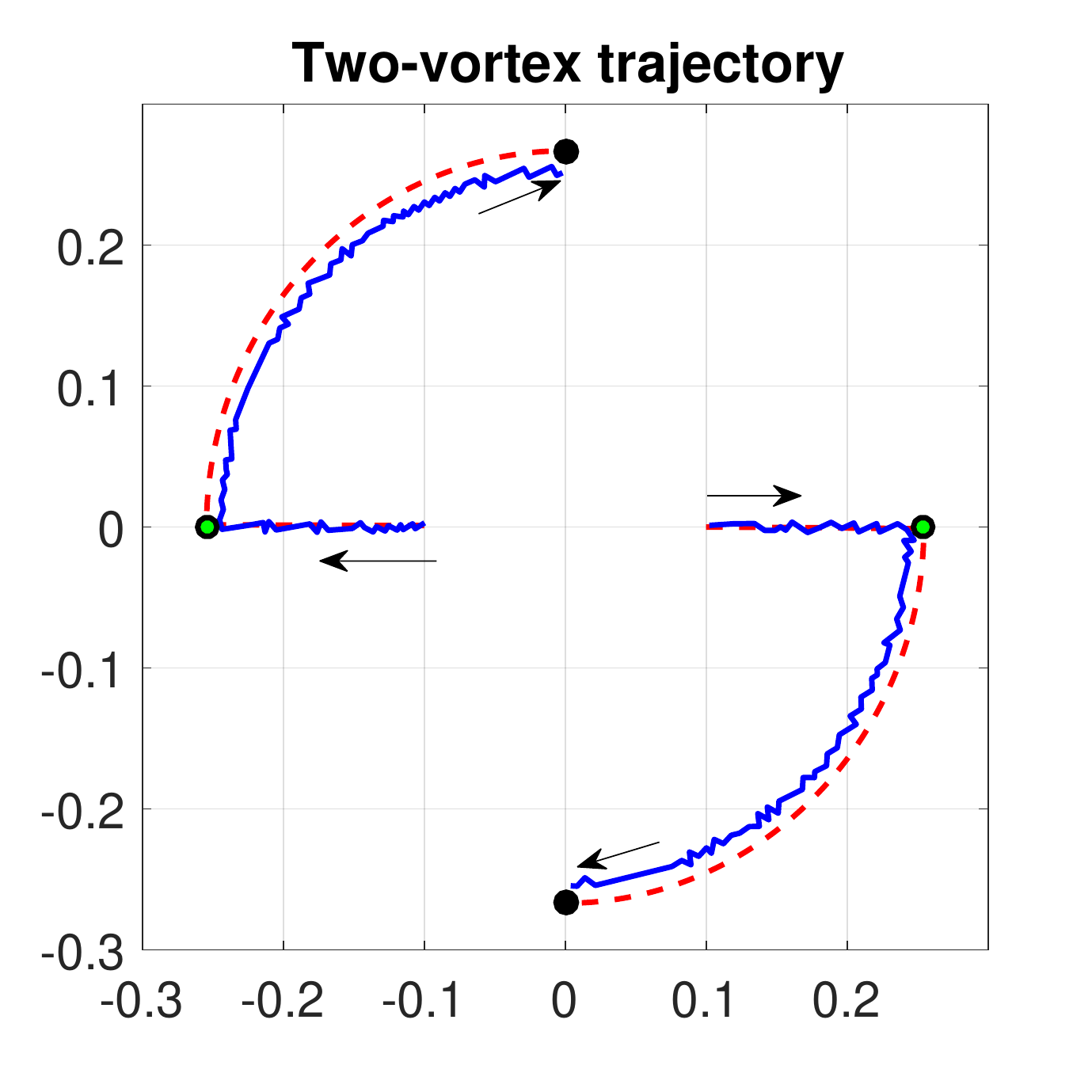}
\caption{}
\label{fig:t2}
\end{subfigure}\caption{(a) $\Omega_{2}$ (the critical rotation
  rate beyond which dipole solutions can be
  identified in an anisotropic trap) as a function of
  anisotropy parameter $b:$ comparison of
full numerics and asymptotics. Dots are obtained from the full numerical
computations of the  PDE (\ref{gp}). The solid line denotes the asymptotic
formula (\ref{Omega2}). The dashed line is the formula (\ref{Omega2A}) derived in
\cite{aftalion2001vortices}. Parameter values are $\gamma=1,~\kappa
=0,~\varepsilon=0.025.$ (b)\ Two-vortex trajectory. Parameter values are
$\varepsilon=0.025,\ \Omega=22.133$ and $b=0.9535.$ Initial conditions consist
of two vortices along x-axis. The arrows indicate the direction of motion. At
first, the vortices approach a saddle point along the x-axis (indicated by
green-black dots). But eventually the two vortices settle along the $y-$axis
(indicated by black dots). Solid curve shows vortex centers from the full PDE
simulation of (\ref{gp}) with $\gamma=1,\kappa=0.$ Dashed line shows the
simulation of the reduced ODE (\ref{dipole}). }%
\label{fig:t}%
\end{figure}

In the pioneering work \cite{aftalion2001vortices}, Aftalion and Du derived a
slightly different formula for $\Omega_{2}$, using a related energy method,
see formula (22) there. Written in our notation, the formula in
\cite{aftalion2001vortices} reads:%
\begin{equation}
\Omega_{2,\text{Aftalion/Du}}=\frac{1+b^{2}}{\nu}+\frac{1+b^{2}}{2}\log\left(
\frac{1+b^{2}}{\nu}\right)  . \label{Omega2A}%
\end{equation}
While both formulae have the same leading-order behaviour in $\nu,$ they have
very different (and large) correction terms. Figure \ref{fig:t}(a)\ shows a
direct comparison between (\ref{Omega2}), (\ref{Omega2A}) and the full
numerical simulations of the PDE (\ref{gp}). Formula (\ref{Omega2})\ appears
to be a significant improvement over (\ref{Omega2A}).

For $\Omega>\Omega_{2},$ the only \emph{potentially stable} solution is the
one corresponding to $r_{-}$ as can be seen by considering perturbations along
the $x-$axis. However this does not tell the whole story:\ a solution may
exist and be stable along the x-axis, but be unstable with respect to the full
spectrum of two-dimensional perturbations. To describe the full stability, as
in section \ref{S:5}, we will -- for simplicity -- consider the overdamed
system $\kappa=0,\gamma=1$ (it can be shown that stability properties are
independent of $\kappa$ as long as $\gamma>0$). The full equations then become%
\begin{align}
\frac{dx_{1}}{dt}  &  =\left(  -2\hat{\Omega}+\frac{2}{1-x_{1}^{2}-b^{2}%
y_{1}^{2}}\right)  x_{1}+\frac{2\nu(x_{1}-x_{2})}{(x_{1}-x_{2})^{2}%
+(y_{1}-y_{2})^{2}}\frac{1-x_{1}^{2}-b^{2}y_{1}^{2}}{1-x_{2}^{2}-b^{2}%
y_{2}^{2}}\nonumber\label{dipole}\\
\frac{dy_{1}}{dt}  &  =\left(  -2\hat{\Omega}+\frac{2}{1-x_{1}^{2}-b^{2}%
y_{1}^{2}}\right)  b^{2}y_{1}+\frac{2\nu(y_{1}-y_{2})}{(x_{1}-x_{2}%
)^{2}+(y_{1}-y_{2})^{2}}\frac{1-x_{1}^{2}-b^{2}y_{1}^{2}}{1-x_{2}^{2}%
-b^{2}y_{2}^{2}}\nonumber\\
\frac{dx_{2}}{dt}  &  =\left(  -2\hat{\Omega}+\frac{2}{1-x_{2}^{2}-b^{2}%
y_{2}^{2}}\right)  x_{2}+\frac{2\nu(x_{2}-x_{1})}{(x_{1}-x_{2})^{2}%
+(y_{1}-y_{2})^{2}}\frac{1-x_{2}^{2}-b^{2}y_{2}^{2}}{1-x_{1}^{2}-b^{2}%
y_{1}^{2}}\\
\frac{dy_{2}}{dt}  &  =\left(  -2\hat{\Omega}+\frac{2}{1-x_{2}^{2}-b^{2}%
y_{2}^{2}}\right)  b^{2}y_{2}+\frac{2\nu(y_{2}-y_{1})}{(x_{1}-x_{2}%
)^{2}+(y_{1}-y_{2})^{2}}\frac{1-x_{2}^{2}-b^{2}y_{2}^{2}}{1-x_{1}^{2}%
-b^{2}y_{1}^{2}}\nonumber
\end{align}
where we defined%
\begin{equation}
\hat{\Omega}:=\frac{\nu\Omega}{1+b^{2}}.
\end{equation}
Linearizing around the equilibrium $x_{1}=r,x_{2}=-r,~y_{1}=y_{2}=0$, we
obtain the following Jacobian matrix,
\[
\left(
\begin{array}
[c]{cccc}%
M_{1} & 0 & M_{3} & 0\\
0 & M_{2} & 0 & M_{4}\\
M_{3} & 0 & M_{1} & 0\\
0 & M_{4} & 0 & M_{2}%
\end{array}
\right)
\]
where%
\begin{align*}
M_{1}  &  =-2\hat{\Omega}+\frac{2}{1-r^{2}}+\frac{4r^{2}}{\left(
1-r^{2}\right)  ^{2}}-\frac{\nu}{2r^{2}}-\frac{2\nu}{1-r^{2}},~~~~ & M_{2}  &
=-2\hat{\Omega}b^{2}+\frac{2b^{2}}{1-r^{2}}+\frac{\nu}{2r^{2}}\\
M_{3}  &  =\frac{\nu}{2r^{2}}-\frac{2\nu}{1-r^{2}},~~~~~~ & M_{4}  &
=-\frac{2\nu}{r^{2}}%
\end{align*}

The eigenvalues of this matrix are easily computed as $M_{1}\pm M_{3}$ and
$M_{2}\pm M_{4}$ which yields,%
\begin{align*}
\lambda_{1}  &  =M_{1}+M_{3}=-2\hat{\Omega}+\frac{2}{1-r^{2}}+\frac{4r^{2}%
}{\left(  1-r^{2}\right)  ^{2}}-\frac{4\nu}{1-r^{2}}~~~~~~ & \lambda_{2}  &
=M_{1}-M_{3}=-2\hat{\Omega}+\frac{2}{1-r^{2}}+\frac{4r^{2}}{\left(
1-r^{2}\right)  ^{2}}-\frac{2\nu}{r^{2}}\\
\lambda_{3}  &  =M_{2}+M_{4}=-2\hat{\Omega}b^{2}+\frac{2b^{2}}{1-r^{2}},~~~~~~
& \lambda_{4}  &  =M_{2}-M_{4}=-2\hat{\Omega}b^{2}+\frac{2b^{2}}{1-r^{2}%
}+\frac{\nu}{r^{2}}.
\end{align*}
Using the relationships $\hat{\Omega}=\frac{\nu}{2r^{2}}+\frac{1}{1-r^{2}}$
and $\Omega>\Omega_{2},$ basic algebra shows that $\lambda_{1,2,3}\leq0.$ On
the other hand, $\lambda_{4}$ becomes%
\[
\lambda_{4}=2(-b^{2}+1)\frac{\nu}{r^{2}}%
\]
and goes through zero precisely at $b=1$; it is stable for $b>1$ and unstable
for $0<b<1.$ The underlying elliptic trap has the form $x^{2}+b^{2}y^{2}=1$.
When $b>1$, the x-axis is the major axis and the y-axis is the minor axis of
the ellipse; the opposite is true for $b<1.$ This shows that the two-vortex
configuration is stable only along the \emph{major} axis.

Figure \ref{fig:t}(b) illustrates this stability result. There, we took
$b=0.9535$, so that the trap is nearly circular but with the extent of the
condensate along the $y-$axis being slightly longer. So we expect a two-vortex
equilibrium to be unstable along the $x-$axis but stable along the $y-$axis.
This is indeed what happens. We ran the imaginary-time integration $\left(
\kappa=0\right)  $ for the full PDE\ (\ref{gp}), starting with initial
conditions consisting of two vortices along the $x-$axis. At first, the two
vortices approach the unstable equilibrium along the $x$-axis (although
unstable, it is a saddle point and initial conditions are along its stable
manifold). However eventually, since this equilibrium is unstable, they travel
towards a stable equilibrium along the $y-$axis.

\section{Large $N$ limit with strongly anisotropic trap}

\label{Sec:5}

\begin{figure}[ptb]
\centering \includegraphics[width=\textwidth]{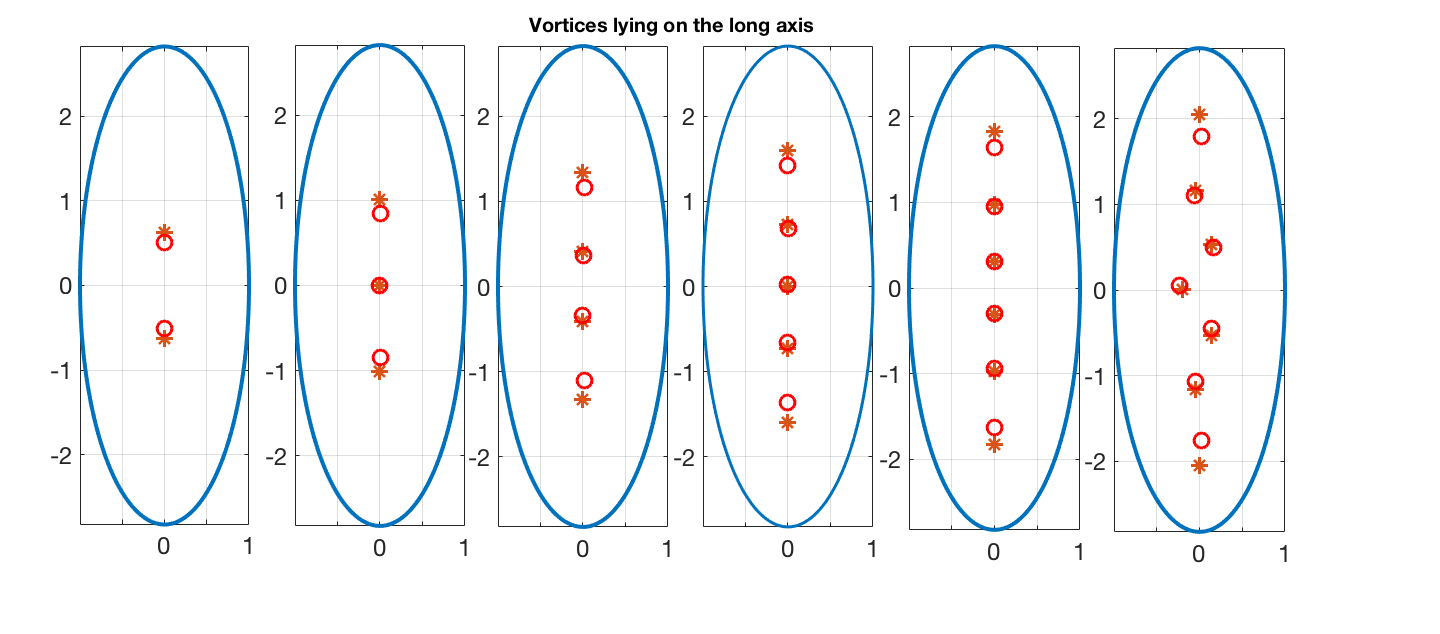}\caption{Comparison
of the steady state of PDE and ODE simulations for $N=2\ldots7$ vortices.
`$\ast$' denotes the steady state of the ODE system (\ref{reduced}) whereas
`o' is from the PDE system (\ref{gp}). The parameters are chosen as:
$\gamma=1,~\kappa=0~,b=\frac{1}{\sqrt{8}},~\varepsilon=0.025$, $~\Omega=6.72 $
for two vortices and $\Omega=7.21$ for three to seven vortices. The boundary
of the elliptical trap $x^{2}+b^{2}y^{2}=1$ is also shown.}%
\label{fig:anisotropy}%
\end{figure}

\label{S:6} We now consider the strongly anisotropic parabolic potential case
of small~\footnote{Notice that for large $b,$ the width of the ellipse
$x^{2}+b^{2}y^{2}=1$ is of $O(1/b).$ Since the size of the vortex core is of
$O(\varepsilon)$, asymptotics require that $b\ll O(1/\varepsilon)$ (otherwise
the vortex size is comparable to the domain size, in which case asymptotics
break down). For this reason, we take the limit $b\rightarrow0$ (high
anisotropy along the y-axis) rather than $b\rightarrow\infty$ (high anisotropy
along the x-axis).} $b$. Figure \ref{fig:anisotropy} illustrates this case with
$b=\frac{1}{\sqrt{8}}.$ For sufficiently strong anisotropy, the vortices align
along the major axis of the elliptic trap (the y-axis in the case
$b\rightarrow0$); see, e.g., also the work of~\cite{jan} for oppositely
charged vortices. Exactly \emph{how} strong depends on the number of vortices
and the exact dependence is an open question that we leave for future study.
For now, we simply assume that the anisotropy is sufficiently strong for the
full alignment to occur, so that the steady state is effectively
one-dimensional. In this case, the ODE system (\ref{reduced})\ reduces motion
purely along the $y-$axis, leading to the following dynamical system of $N$
variables:\bes\label{ode1d}%
\begin{equation}
y_{jt}=\left(  -2\hat{\Omega}+\frac{2}{1-b^{2}y_{j}^{2}}\right)  y_{j}%
+2\nu\sum_{k\neq j}\frac{1-b^{2}y_{j}^{2}}{1-b^{2}y_{k}^{2}}\frac{y_{j}-y_{k}%
}{|y_{j}-y_{k}|^{2}},\label{2:00}%
\end{equation}
where%
\begin{equation}
\hat{\Omega}:=\nu\frac{\Omega}{1+b^{2}}.
\end{equation}
\ees(where for simplicity we took the overdamped limit $\gamma=1,\kappa=0$).
Define $z_{j}=by_{j}$ so that (\ref{2:00}) becomes%
\[
\frac{1}{b^{2}}z_{jt}=\left(  -2\hat{\Omega}+\frac{2}{1-z_{j}^{2}}\right)
z_{j}+2\nu\sum_{k\neq j}\frac{1-z_{j}^{2}}{1-z_{k}^{2}}\frac{z_{j}-z_{k}%
}{|z_{j}-z_{k}|^{2}}.
\]
We wish to compute the effective one-dimensional density of the resulting
steady state in the continuum limit $N\rightarrow\infty$ of this system. As in
\S \ref{S:5}, we define the one-dimensional density to be%
\[
\rho(z)=\sum\delta(z-z_{j}).
\]
The steady-state density then satisfies\bes\label{4:10}%
\begin{equation}
\left(  -\hat{\Omega}+\frac{1}{1-z^{2}}\right)  z+\nu\left(  1-z^{2}\right)
\mathchoice {{\setbox0=\hbox{$\displaystyle{\textstyle -}{\int}$} \vcenter{\hbox{$\textstyle
-$}}\kern-.5\wd0}}{{\setbox0=\hbox{$\textstyle{\scriptstyle -}{\int}$} \vcenter{\hbox{$\scriptstyle
-$}}\kern-.5\wd0}}{{\setbox0=\hbox{$\scriptstyle{\scriptscriptstyle
-}{\int}$} \vcenter{\hbox{$\scriptscriptstyle
-$}}\kern-.5\wd0}}{{\setbox0=\hbox{$\scriptscriptstyle{\scriptscriptstyle
-}{\int}$} \vcenter{\hbox{$\scriptscriptstyle -$}}\kern-.5\wd0}}\!\int
_{-a}^{a}\frac{1}{y-z}\frac{1}{1-y^{2}}\rho(y)dy=0\label{12:17}%
\end{equation}
where
$\mathchoice {{\setbox0=\hbox{$\displaystyle{\textstyle -}{\int}$} \vcenter{\hbox{$\textstyle
-$}}\kern-.5\wd0}}{{\setbox0=\hbox{$\textstyle{\scriptstyle -}{\int}$} \vcenter{\hbox{$\scriptstyle
-$}}\kern-.5\wd0}}{{\setbox0=\hbox{$\scriptstyle{\scriptscriptstyle
-}{\int}$} \vcenter{\hbox{$\scriptscriptstyle
-$}}\kern-.5\wd0}}{{\setbox0=\hbox{$\scriptscriptstyle{\scriptscriptstyle
-}{\int}$} \vcenter{\hbox{$\scriptscriptstyle -$}}\kern-.5\wd0}}\!\int
_{-a}^{a}$ denotes the Cauchy principal value integral. Here, $a$ is the
radius of the one-dimensional vortex \textquotedblleft lattice\textquotedblright. The solution
to (\ref{12:17})\ is subject to the additional mass constraint%
\begin{equation}
\int_{-a}^{a}\rho(z)dz=N
\end{equation}
\ees Together, equations (\ref{4:10})\ are to be solved for both the density
$\rho(z)$ and the radius $a.$

\begin{figure}[tb]
\centering\begin{subfigure}[b]{0.31\textwidth}
\includegraphics[width=\textwidth]{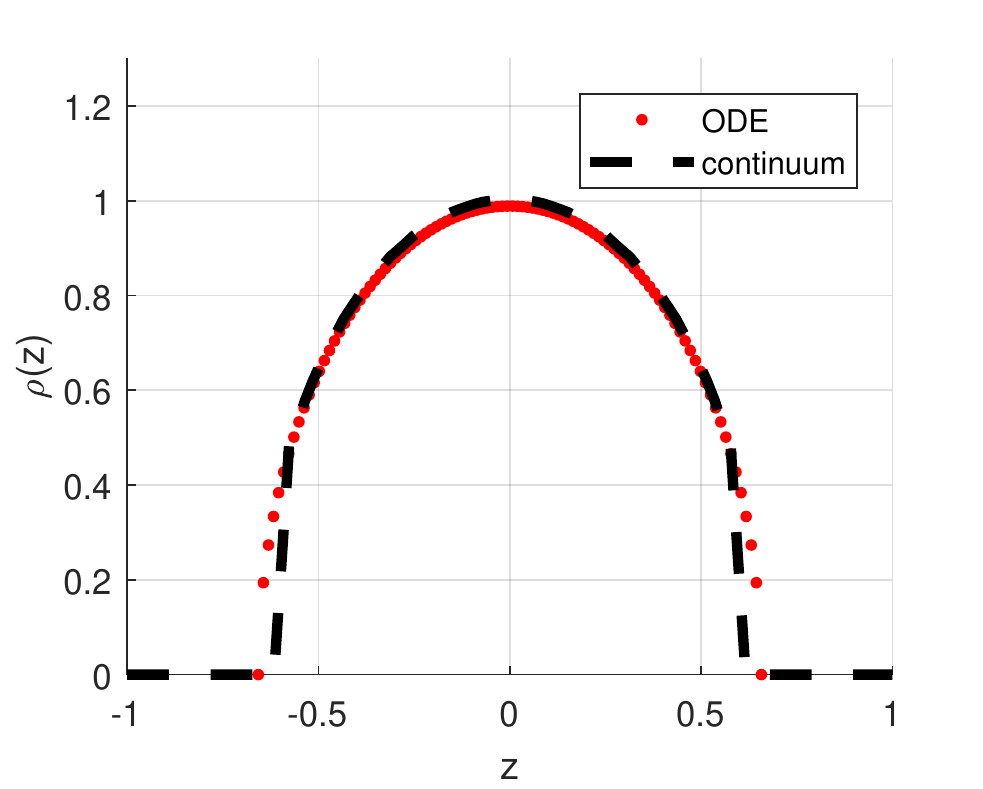}
\caption{}
\label{fig:t3}
\end{subfigure}\begin{subfigure}[b]{0.31\textwidth}
\includegraphics[width=\textwidth]{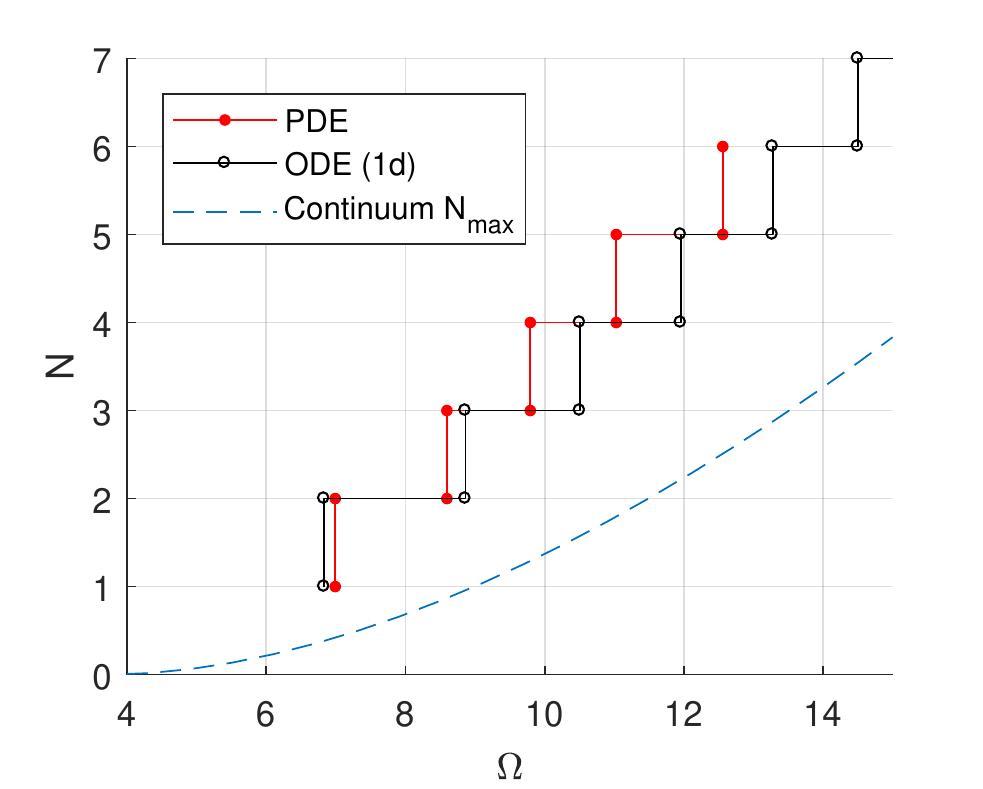}
\caption{}
\end{subfigure}\begin{subfigure}[b]{0.31\textwidth}
\includegraphics[width=\textwidth]{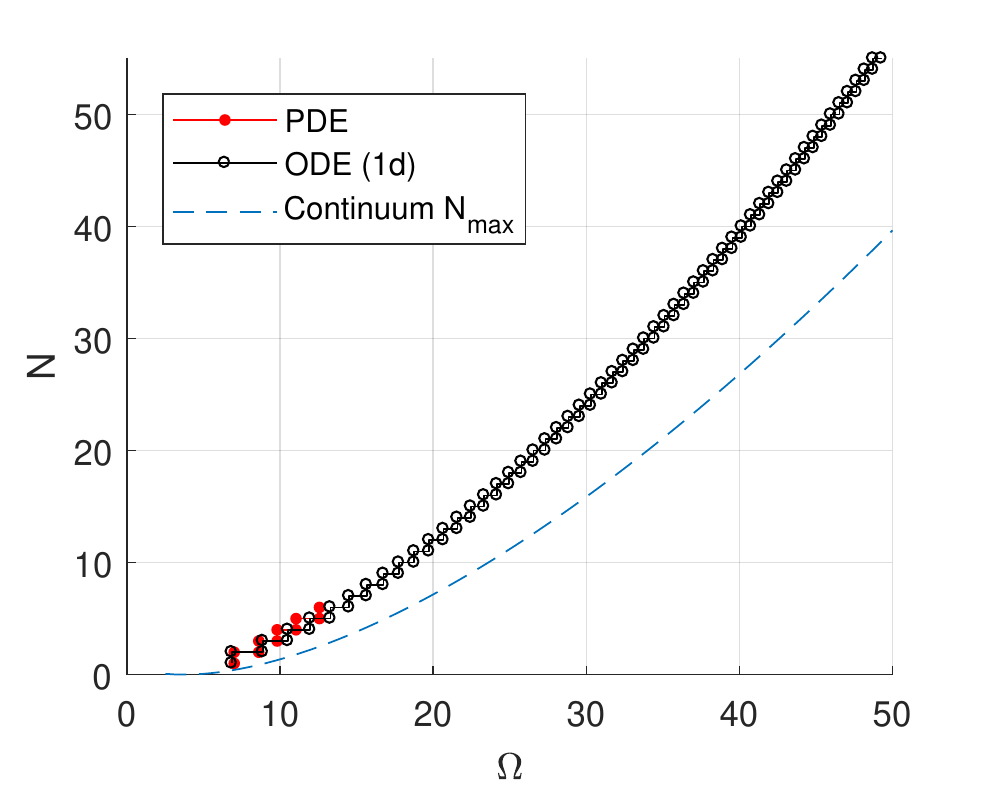}
\caption{}
\label{fig:t4}
\end{subfigure}
\caption{ (a) Steady state density of the ODE system (\ref{ode1d}), compared
with the continuum limit (\ref{rho1d}), where only eight terms of the series
is used. Here, $N=40$ and $\hat{\Omega}=40$ (b) Maximal admissible number of
vortices for the full PDE simulation of (\ref{gp}) versus the ODE system
(\ref{ode1d}), versus the versus continuum formula (\ref{NNmax}). 
Parameters are $\gamma=1,
\kappa=0, \varepsilon=0.025$, $b=\frac{1}{\sqrt{8}}$ and $\Omega$ is slowly decreasing
according to the formula $\Omega=10-10^{-4}t$. (c) Comparison of the ODE
(\ref{ode1d}) and continuum limit formula (\ref{NNmax}) with ODE motion
restricted to the y-axis, for larger number of vortices. Same parameters as in
(b), except that $\Omega=60-10^{-4}t$. }%
\label{fig:t5}%
\end{figure}

A solution to (\ref{4:10})\ can be derived using techniques involving the
Chebychev polynomials, as suggested by \cite{shestopalov2002integral}, see
Chapter 18 there (the Fourier--Chebyshev series). We start by recalling the
following standard identities between Chebyshev polynomials $U_{n}$ and $T_{n}$:\bes\label{1:23}%
\begin{align}
\mathchoice {{\setbox0=\hbox{$\displaystyle{\textstyle -}{\int}$} \vcenter{\hbox{$\textstyle -$}}\kern-.5\wd0}}{{\setbox0=\hbox{$\textstyle{\scriptstyle -}{\int}$} \vcenter{\hbox{$\scriptstyle -$}}\kern-.5\wd0}}{{\setbox0=\hbox{$\scriptstyle{\scriptscriptstyle -}{\int}$} \vcenter{\hbox{$\scriptscriptstyle -$}}\kern-.5\wd0}}{{\setbox0=\hbox{$\scriptscriptstyle{\scriptscriptstyle -}{\int}$} \vcenter{\hbox{$\scriptscriptstyle -$}}\kern-.5\wd0}}\!\int
_{-1}^{1}\frac{\sqrt{1-y^{2}}U_{n-1}(x)}{y-x}dy  &  =-\pi T_{n}%
(x)\label{1:23a}\\
\mathchoice {{\setbox0=\hbox{$\displaystyle{\textstyle -}{\int}$} \vcenter{\hbox{$\textstyle -$}}\kern-.5\wd0}}{{\setbox0=\hbox{$\textstyle{\scriptstyle -}{\int}$} \vcenter{\hbox{$\scriptstyle -$}}\kern-.5\wd0}}{{\setbox0=\hbox{$\scriptstyle{\scriptscriptstyle -}{\int}$} \vcenter{\hbox{$\scriptscriptstyle -$}}\kern-.5\wd0}}{{\setbox0=\hbox{$\scriptscriptstyle{\scriptscriptstyle -}{\int}$} \vcenter{\hbox{$\scriptscriptstyle -$}}\kern-.5\wd0}}\!\int
_{-1}^{1}\frac{T_{n(x)}}{(y-x)\sqrt{1-y^{2}}}dy  &  =\pi U_{n-1}%
(x)\label{1:23b}\\
\mathchoice {{\setbox0=\hbox{$\displaystyle{\textstyle -}{\int}$} \vcenter{\hbox{$\textstyle -$}}\kern-.5\wd0}}{{\setbox0=\hbox{$\textstyle{\scriptstyle -}{\int}$} \vcenter{\hbox{$\scriptstyle -$}}\kern-.5\wd0}}{{\setbox0=\hbox{$\scriptstyle{\scriptscriptstyle -}{\int}$} \vcenter{\hbox{$\scriptscriptstyle -$}}\kern-.5\wd0}}{{\setbox0=\hbox{$\scriptscriptstyle{\scriptscriptstyle -}{\int}$} \vcenter{\hbox{$\scriptscriptstyle -$}}\kern-.5\wd0}}\!\int
_{-1}^{1}\frac{T_{n}(x)T_{m}(x)}{\sqrt{1-y^{2}}}dy  &  =\left\{
\begin{array}
[c]{ll}%
0~~ & n\neq m\\
\pi~~ & n=m=0\\
\pi/2 & n=m\neq0
\end{array}
\right. \label{1:23c}\\
\mathchoice {{\setbox0=\hbox{$\displaystyle{\textstyle -}{\int}$} \vcenter{\hbox{$\textstyle -$}}\kern-.5\wd0}}{{\setbox0=\hbox{$\textstyle{\scriptstyle -}{\int}$} \vcenter{\hbox{$\scriptstyle -$}}\kern-.5\wd0}}{{\setbox0=\hbox{$\scriptstyle{\scriptscriptstyle -}{\int}$} \vcenter{\hbox{$\scriptscriptstyle -$}}\kern-.5\wd0}}{{\setbox0=\hbox{$\scriptscriptstyle{\scriptscriptstyle -}{\int}$} \vcenter{\hbox{$\scriptscriptstyle -$}}\kern-.5\wd0}}\!\int
_{-1}^{1}U_{n}(x)U_{m}(x)\sqrt{1-y^{2}}dy  &  =\left\{
\begin{array}
[c]{ll}%
0~~ & n\neq m\\
\pi/2~~ & n=m=0
\end{array}
\right.  . \label{1:23d}%
\end{align}
\ees Identity (\ref{1:23a}) as well as the form of the integral
equation\ (\ref{12:17})\ motivates the following anzatz for the density
$\rho:$ \bes\label{rho1d}%
\begin{equation}
\rho(z)=-\frac{1}{\pi}\sum_{i=1}^{\infty}c_{i}U_{i-1}(\frac{z}{a})\left(
1-z^{2}\right)  \sqrt{1-\frac{z^{2}}{a^{2}}}. \label{1:25}%
\end{equation}
Using (\ref{1:23d})\ in Eq.~(\ref{12:17}) then yields the following expression
for $c_{i}$ in terms of $a:$%
\begin{equation}
c_{i}=\frac{2}{\pi}\int_{-1}^{1}\left(  -\hat{\Omega}+\frac{1}{1-a^{2}y^{2}%
}\right)  \frac{ay}{\nu\left(  1-a^{2}y^{2}\right)  }T_{i}(y)\frac{1}%
{\sqrt{1-y^{2}}}dy. \label{1:26}%
\end{equation}
Upon substituting (\ref{1:25})\ into (\ref{12:17}) and using identities
(\ref{1:23}) we obtain
\begin{equation}
\int_{-a}^{a}\rho(z)dz=-\frac{a}{2}\left(  c_{1}(1-\frac{a^{2}}{4}%
)-\frac{a^{2}}{4}c_{3}\right)  =N
\end{equation}
Evaluating $c_{1}$ and $c_{3}$ using (\ref{1:26})\ finally yields the
following relationship between $N$ and $a,$
\begin{equation}
N=\frac{1}{\nu}\left(  \frac{\hat{\Omega}a^{2}}{2\sqrt{1-a^{2}}}-\frac
{(a^{2}-2)^{2}}{\nu(1-a^{2})^{\frac{3}{2}}}+1\right)  . \label{Fold_ani}%
\end{equation}
\ees Note that while the expression for the radius $a$ is explicit, the
density $\rho\left(  z\right)  $ itself does not appear to have a closed form
solution, having an infinite-series representation\ (\ref{1:25}).\ However the
coefficients $c_{i}$ in (\ref{1:25}) are easy to compute numerically, while in
practice the series representation converges very quickly. Figure
\ref{fig:t5}(a) shows a direct comparison between the analytical density
(\ref{1:25}) and the steady state of (\ref{4:10})\ with $N=40$, verifying
that the analytical prediction is in very good agreement with the
numerical ODE result.

The function $a\rightarrow N(a)$ has a unique maximum at $a^{2}=2\left(
\hat{\Omega}-1\right)  /(2\hat{\Omega}+1)$, given by%
\begin{equation}
N_{\max,\text{1d}}=\frac{1}{\nu}\left(  1+3^{-3/2}(\hat{\Omega}-4)\sqrt
{1+2\hat{\Omega}}\right)  . \label{NNmax}%
\end{equation}
This provides the asymptotic upper bound for the number of vortices that can
be aligned along the x-axis. This is the main result of this section,
concluding the derivation of (\ref{12:08}). Figure \ref{fig:t5}(c)\ shows the
comparison between the formula (\ref{NNmax}) and the ODE. Although it appears
that the two curves diverge, their ratio approaches $1$ as $\hat{\Omega}$ is
increased; a similar comparison but for small values
of $\Omega$ is shown in Fig.~\ref{fig:t5}(b).

\section{Discussion}

In this paper we derived a novel and more accurate set of ODEs (\ref{reduced}%
) for vortex motion in BEC with an (isotropic, as well as
with an) anisotropic trap. These ODEs incorporate
the effect of the trap inhomogenuity on vortex-to-vortex interactions. In
turn, the analysis of ODEs yields an accurate analytical formula for the
vortex lattice density, as well as the maximal admissible number of vortices
$N_{\max}$ as a function of rotation rate $\Omega$ under two
scenarios:\ isotropic trap with large $N,$ and high-anisotropy regime with
large $N.$ Additionally, we examined existence and stability of two vortices
in an anisotropic trap; i.e., we focused both on the fundamental
building block of the inter-vortex interactions and the
large $N$
``vortex crystal'' limit. For the isotropic case, we used techniques from
swarming literature \cite{kolokolnikov2014tale, fetecau2011swarm} to estimate
the large-$N$ vortex lattice density. In the case of high-anisotropy, we used
Chebychev expansions to explicitly compute the critical thresholds
and analyze the vortex density.

It would be interesting to redo the analysis in \cite{navarro2013dynamics} for
the new ODE system (\ref{reduced}). For example, it would be relevant to
identify in that context the asymmetric
configurations of two vortices, as well as to extend considerations beyond the
case of two, i.e., to triplets of vortices, as well as beyond.

Our results improve upon known results in the literature in two ways. The
reduced system of motion (\ref{reduced}) is more accurate than previously
reported in e.g. \cite{navarro2013dynamics, kolokolnikov2014tale}
(see~\cite{siambook} for a relevant discussion of earlier models). As a
consequence, we have obtained more accurate thresholds for existence and
stability, especially in the case of multiple vortices, but also in the case
of two vortices within an anisotropic trap. Numerical experiments show that
these thresholds improve also upon those found in \cite{aftalion2001vortices},
for example.

\begin{figure}[tb]
{}
\par
\begin{center}
\includegraphics[width=0.5\textwidth]{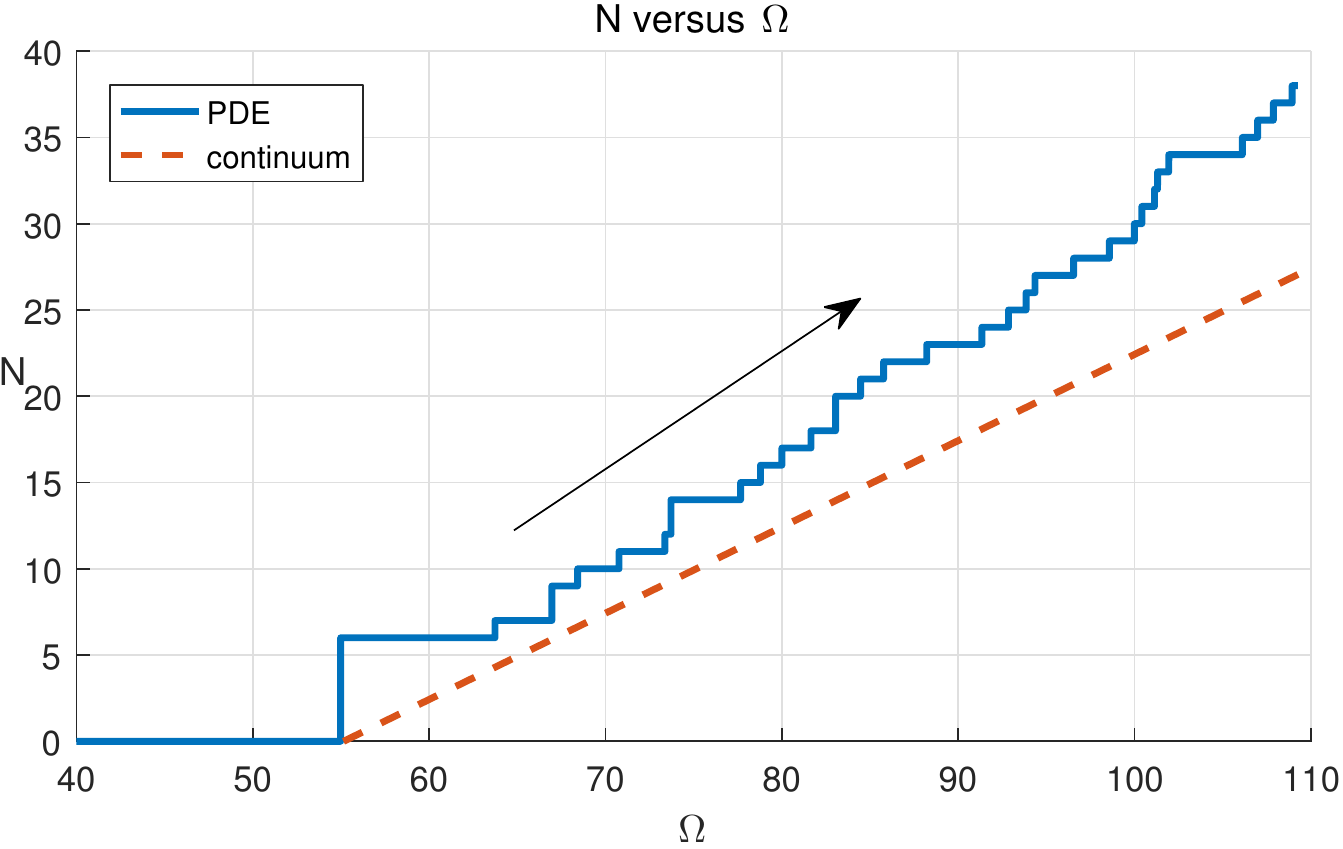}
\end{center}
\par
\caption{Number of vortices as a function of very slowly increasing
$\Omega:\Omega=10^{-2}t.$ All other parameters are as in Figure \ref{fig:Nmax}%
. The curve \textquotedblleft PDE\textquotedblright\ is from PDE simulations
whereas the curve \textquotedblleft continuum\textquotedblright\ is the
asymptotic estimate (\ref{Nmin}). }%
\label{fig:Nmin}%
\end{figure}

It is interesting to note that that in addition to the upper bound $N_{\max}$,
there is also a lower bound on the number of vortices, $N_{\min}$, for a given
rotation rate $\Omega.$ As $\Omega$ is sufficiently increased, vortices
spontaneously nucleate from the Thomas-Fermi boundary. In the case of an
isotropic trap, a zero-vortex state becomes unstable as $\Omega$ increases
past $\underline{\Omega}\sim2.561\varepsilon^{-2/3}$ -- see
\cite{anglin2001local, carretero2016vortex, tzou2016weakly} for derivation.
This computation can be extended to a single vortex at the center of degree
$N.$ In this case, one finds that the stability threshold is $\underline
{\Omega}\sim2.53\varepsilon^{-2/3}+2N.$ Solving for $N$, this in turn yields
the formula%
\begin{equation}
N_{\min}\sim\frac{\Omega}{2}-1.28\varepsilon^{-2/3}. \label{Nmin}%
\end{equation}
Speculatively, let us now make a very crude approximation, and na\"{\i}vely
assume that the entire vortex lattice of $N$ vortices can be approximated by a
single vortex of degree $N$ at the origin. This assumption is clearly
incorrect if the vortex lattice occupies the entire trap, but may be
reasonable if we suppose that the entire vortex lattice is clustered near the
center and away from the Thomas-Fermi boundary. In any case, under this
very crude
assumption, (\ref{Nmin}) provides an asymptotic approximation to the
\emph{lower bound} for existence of $N$ vortices as a function of $\Omega$, so
that $N_{\min}<N<N_{\max}.$ Surprisingly, this actually works relatively well
in practice, at least for relatively small vortex numbers as Figure
\ref{fig:Nmin} illustrates. An open question is to extend this bound to an
anisotropic trap, as well as the situation where the vortex lattice is spread
throughout the trap, and cannot be easily reduced to a single $N-$degree
vortex.

In conclusion, direct asymptotic reduction of the GPE, combined with
coarse-graining techniques for large number of vortices
(and bifurcation analysis for small $N$ vortex clusters) provide a powerful
set of tools that yields novel insights into a well-studied classical problem of Bose-Einstein\ Condensates.

\bibliographystyle{elsarticle-num}
\bibliography{elliptic_v2}

\end{document}